\documentclass[aps,pre,reprint,showpacs,dvipdfmx]{revtex4-1}
\usepackage{graphicx}
\usepackage{dcolumn}
\usepackage{amsmath}
\usepackage{amssymb}
\usepackage{color}
\usepackage{subfigure}
\usepackage{bm}
\usepackage{here}
\usepackage{tabularx}
\newcolumntype{C}{>{\centering\arraybackslash}X}
\newcolumntype{L}{>{\raggedright\arraybackslash}X}
\newcolumntype{R}{>{\raggedleft\arraybackslash}X}

\newcommand{\modHN}[1]{{\color{red} #1}}
\usepackage{comment}

\begin{document}
\preprint{HEP/123-qed}
\title{Equilibrium measurement method of slip length based on fluctuating hydrodynamics}
\author{Hiroyoshi Nakano}
\affiliation{Department of Physics, Kyoto University,
Kyoto 606-8502, Japan}
\author{Shin-ichi Sasa}
\affiliation{Department of Physics, Kyoto University,
Kyoto 606-8502, Japan}
\date{\today}

\begin{abstract}
We perform equilibrium molecular dynamics simulations for nanoscale fluids confined between two parallel walls and investigate how the autocorrelation function of force acting on one wall is related to the slip length. We demonstrate that for atomically smooth surfaces, the autocorrelation function is accurately described by linearized fluctuating hydrodynamics (LFH). Excellent agreement between the simulation and the LFH solution is found over a wide range of scales, specifically, from the time scale of fluid relaxation even to that of molecular motion. Fitting the simulation data yields a reasonable estimation of the slip length. We show that LFH provides a starting point for examining the relationship between the slip length and the force fluctuations.
\end{abstract}


\maketitle
\section{Introduction} 
Over the last two decades, because of remarkable developments in nanotechnology, there has been considerable interest in ``nanofluidics," which involves a quantitative description of fluid motion at the nanoscale ($<100nm$)~\cite{eijkel2005nanofluidics,sparreboom2009principles,bocquet2010nanofluidics}. One of the core concepts is the assumption of partial slip boundary condition; that is, fluid velocity is linearly proportional to the shear rate at fixed solid surfaces~\cite{navier1823,lamb1993hydrodynamics,happel2012low}:
 \begin{eqnarray}
v \Big|_{\rm wall} = b \frac{\partial v}{\partial z}\Big|_{\rm wall},
\label{eq:partial slip boundary condition, introduction}
\end{eqnarray}
where $v$ is the tangential velocity relative to the surface and $z$ is the coordinate along the surface normal. $b$ is the slip length, which characterizes the extent of boundary slip. It has been confirmed in recent experiments that a typical slip length is about $0-30nm$ for smooth surfaces; the effect of boundary slip becomes non-negligible as the size of the channel confining the fluid approaches nano- and micro- scales. Therefore, aiming for control and manipulation of nanoscale fluids, much effort has been devoted to the investigation of factors that affect the slip length~\cite{vinogradova1999slippage,neto2005boundary,lauga2007microfluidics,cao2009molecular,lee2014interfacial}.

Bocquet and Barrat developed a useful method to examine how the slip length depends on the microscopic structure of the solid surface~\cite{bocquet1994hydrodynamic}. 
\modHN{They focused on another expression of the partial slip boundary condition; specifically, the total force $F_{\rm wall}$ acting on the wall is linearly proportional to the slip velocity $v |_{\rm wall}$ at the wall~\cite{navier1823}:
\begin{eqnarray}
F_{\rm wall} = - S \lambda v \Big|_{\rm wall},
\label{eq:def of the friction coefficient}
\end{eqnarray}
where $S$ is the surface area and $\lambda$ the liquid-solid friction coefficient. As initially discussed by Navier~\cite{navier1823}, by assuming that the bulk constitutive equation holds adjacent to the wall:
\begin{eqnarray}
F_{\rm wall} = - \eta \frac{\partial v}{\partial z}\Big|_{\rm wall},
\label{eq:assumption of the bulk constitutive equation in intro}
\end{eqnarray} 
where $\eta$ is the viscosity of the bulk, the friction coefficient is connected with the slip length as
\begin{eqnarray}
b = \frac{\eta}{\lambda}.
\end{eqnarray}
Based on the non-equilibrium statistical mechanics, Bocquet and Barrat proposed the following expressions for the friction coefficient and the slip length:
\begin{eqnarray}
\lambda = \gamma(\tau_0) ,
\label{eq: expression of slip length: bocquet and barrat}\\[3pt]
b = \frac{\eta}{\gamma(\tau_0)}
\label{eq: expression of slip length2: bocquet and barrat}
\end{eqnarray}
with
\begin{eqnarray}
\gamma(\tau) = \frac{1}{Sk_B T}\int_0^{\tau} dt \langle \hat{F}(t) \hat{F}(0) \rangle_{\rm eq},
\end{eqnarray}
where $k_B$ the Boltzmann constant, $T$ the temperature of the fluid, $\langle \cdot \rangle_{\rm eq}$ the canonical ensemble average at temperature $T$, and $\hat{F}(t)$ the total microscopic force between the wall and fluid at time $t$. $\tau_0$ is taken to the first zero of $\langle \hat{F}(t) \hat{F}(0) \rangle_{\rm eq}$. 
}

The relations (\ref{eq: expression of slip length: bocquet and barrat}) and (\ref{eq: expression of slip length2: bocquet and barrat}) were mainly applied to atomically smooth surfaces and successfully extracted some general relationship between the friction coefficient (or the slip length) and the microscopic parameters of walls. For example, the relation (\ref{eq: expression of slip length2: bocquet and barrat}) explains how the slip length depends on microscopic parameters such as the interaction between a fluid particle and a solid particle~\cite{barrat1999influence,huang2008water} and the curvature of the wall surface~\cite{falk2010molecular}. Also, the quantitative relationship between the slip length and the static properties of the fluid adjacent to the wall, such as the density and structure factor, is derived from the relation (\ref{eq: expression of slip length2: bocquet and barrat})~\cite{priezjev2004molecular,priezjev2006influence}.

However, the theoretical interpretation of the relations (\ref{eq: expression of slip length: bocquet and barrat}) and (\ref{eq: expression of slip length2: bocquet and barrat}) is not completely understood, and it is unclear why they are related to the friction coefficient and slip length~\cite{petravic2007equilibrium,huang2014green,ramos2016hydrodynamic}. Hence, their application has been limited to atomically smooth surfaces. To elucidate whether they can provide the starting point to examine the boundary conditions for more realistic and complicated surfaces, it is important to clarify their theoretical interpretation.

One approach to this problem is to connect the relation (\ref{eq: expression of slip length: bocquet and barrat}) with the Green--Kubo formula, which is formally expressed in the form~\cite{bocquet1994hydrodynamic,fuchs2002statistical,kobryn2008molecular,bocquet2013green,chen2015determining,camargo2019boundary}
\begin{eqnarray}
\lambda = \lim_{\tau \to \infty}\lim_{L \to \infty} \gamma(\tau).
\label{eq: expression of slip length: bocquet and barrat, infty}
\end{eqnarray}
The long-time limit $\tau \to \infty$ and thermodynamic limit $L\to \infty$ are essential to obtain the correct friction coefficient from the Green--Kubo formula. The linear response theory explains the importance of these limits as follows~\cite{nakano2019statistical,kirkwood1946statistical,zubarev1996statistical}. There are two characteristic time scales in a confined fluid; one is that of the microscopic motion of molecules (denoted by $\tau_{\rm micro}$) and the other is that of the global equilibration of fluid (denoted by $\tau_{\rm macro}$). When the system size is sufficiently large, $\tau_{\rm micro} \ll \tau_{\rm macro}$ holds because the relaxation time of slow variables is larger for larger system size. Then, because $\hat{F}(\tau)$ is a fast variable, it is reasonable to assume that $\langle  \hat{F}(\tau) \hat{F}(0)\rangle_{\rm eq}$ decays to $0$ at a sufficiently short time relative to $\tau_{\rm macro}$. However, $\langle  \hat{F}(\tau) \hat{F}(0)\rangle_{\rm eq}$ does not completely decay to zero even at $\tau\simeq \tau_{\rm macro}$ because fast variables are generally coupled with slow variables at the $\tau_{\rm macro}$-scale. Actually, for finite size systems $\gamma(\tau)$ approaches a geometry-dependent value in the long time limit $\tau \gg \tau_{\rm macro}$~\cite{bocquet1994hydrodynamic,petravic2007equilibrium}. Therefore, to obtain the slip length from $\gamma(\tau)$, it is necessary to remove the contributions of the slow variables from the time integral of $\gamma(\tau)$. This is accomplished by taking the thermodynamic limit $L\to \infty$ before the long-time limit $\tau \to \infty$, because the dynamics of the slow variables is separated in the limit $L\to \infty$.
When the slow and fast variables are completely separated, it is expected that $\langle  \hat{F}(\tau) \hat{F}(0)\rangle_{\rm eq}$ would be equal to $0$ in the time region $\tau_{\rm micro} \ll \tau \ll \tau_{\rm macro}$; as a result, $\gamma(\tau)$ has a well-defined plateau region where $\gamma(\tau)$ is almost constant. The value of $\gamma(\tau)$ in this plateau region is the correct friction coefficient.

Thus, in order to connect the relation (\ref{eq: expression of slip length: bocquet and barrat}) and the Green--Kubo formula, we have to identify $\tau_0$ in the relation (\ref{eq: expression of slip length: bocquet and barrat}) with the plateau region. So far, however, its connection remains unclear for the following reasons; (i) Previous studies could not prove the existence of the well-defined plateau region around $\tau_0$~\cite{petravic2007equilibrium,huang2014green,ramos2016hydrodynamic}; (ii) the relation (\ref{eq: expression of slip length: bocquet and barrat}) holds even for the extremely small system where the assumption $\tau_{\rm micro} \ll \tau_{\rm macro}$ is expected to break down~\cite{bocquet1994hydrodynamic,falk2010molecular,tocci2014friction,wei2014breakdown,liang2015slip}. \modHN{Additionally, recently, J. A. de la Torre \textit{et al.} discussed the behavior of $\gamma(\tau)$ by using the Mori projection operator method and claimed that $\gamma(\tau)$ does not exhibit the plateau region even in the thermodynamic limit and the formal expression of Green--Kubo formula, (\ref{eq: expression of slip length: bocquet and barrat, infty}), is not correct~\cite{PhysRevE.99.022126,PhysRevLett.123.264501}.}

There are two main themes in this paper: (i) to propose a new theoretical interpretation of the relations (\ref{eq: expression of slip length: bocquet and barrat}) and (\ref{eq: expression of slip length2: bocquet and barrat}); (ii) to develop a new equilibrium measurement method. We study them for the atomically smooth walls, where the bulk constitutive equation (\ref{eq:assumption of the bulk constitutive equation in intro}) holds even near the wall. Then, we basically focus on the relation (\ref{eq: expression of slip length2: bocquet and barrat}) and refer to it as BB's relation. Our crucial idea is to analyze the force autocorrelation function in linearized fluctuating hydrodynamics (LFH). Although LFH was originally developed as a phenomenological model to describe the dynamics of macroscopic fluctuations~\cite{landau1959course,zwanzig1961memory,kawasaki1973simple,garcia1991fluctuating,mareschal1992dynamic}, we demonstrate that it reproduces the results of numerical simulations even at the time scale of molecular motion. Furthermore, by fitting the simulation data to the LFH solution, a reasonable estimation of the slip length is obtained. From these results, we find that LFH is a reasonable starting point for examining the relationship between the fluctuations of force acting on walls and the slip length. Then, by combining LFH and the molecular dynamics (MD) simulation, we study the validity of BB's relation.

The remainder of this paper is organized as follows. In Sec.~\ref{sec:Microscopic description of confined fluids}, we introduce a microscopic model of a confined fluid. In Sec.~\ref{subsec:Non-equilibrium molecular dynamics simulation}, we investigate in advance the basic properties of the fluid and wall by using the non-equilibrium molecular dynamics (NEMD) simulation. In Sec.~\ref{sec:Review of BB's formula}, we review BB's relation with the equilibrium molecular dynamics (EMD) simulation and sort out the problem. In Sec.~\ref{sec:universal model for force autocorrelation function}, we introduce LFH and give some exact results derived from it. In Sec.~\ref{sec:Validity of LFH description}, by comparing the EMD simulation result with the LFH solution, we demonstrate that LFH accurately describes the fluctuations of the force acting on the wall. We also propose a new equilibrium measurement method for the slip length. Then, in Sec.~\ref{sec:BB's relation revisited}, through the analysis of LFH, we propose the new theoretical interpretation of BB's relation. Sec.~\ref{sec:Discussion} is \modHN{devoted} to conclusions and discussions.

\section{Microscopic description of confined fluids}
\label{sec:Microscopic description of confined fluids}

\subsection{Model}
\label{subsec:model}
\begin{figure}[htbp]
\centering
\includegraphics[width=4cm]{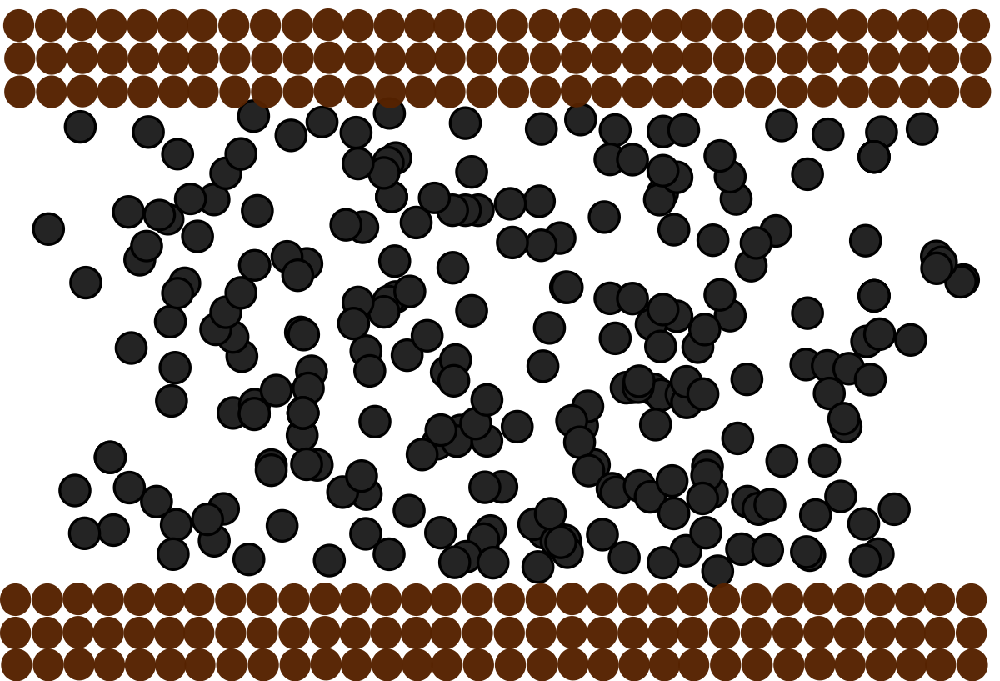}
\hspace{0.3cm}
\includegraphics[width=4cm]{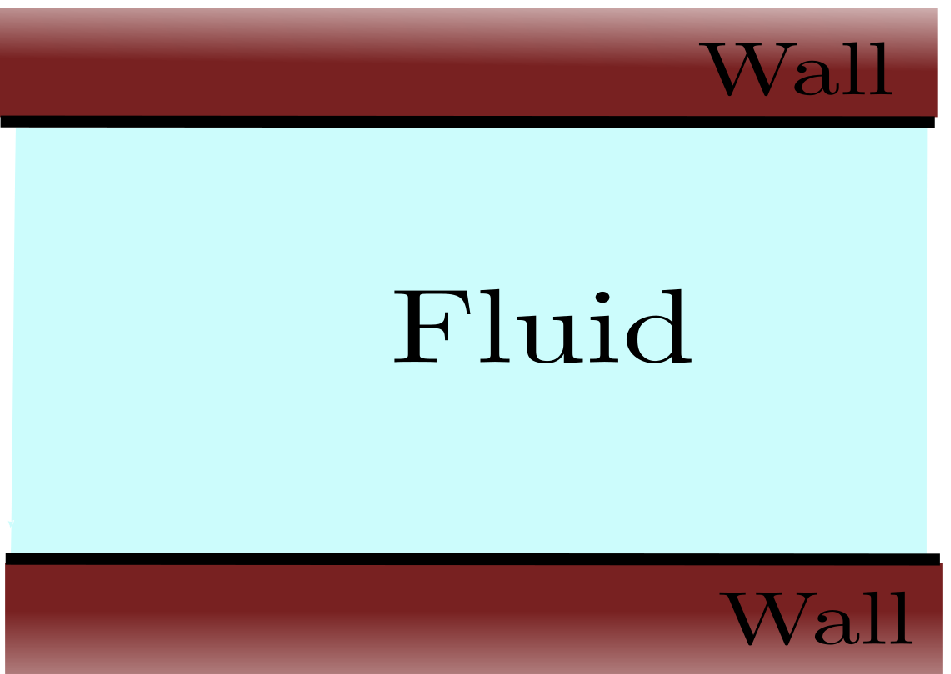} 
\caption{(Color online) Schematic illustration of the model. on the left is a microscopic description of the molecules, and on the right is a macroscopic description of the fluid.}
\label{fig:schematic illustration of the model}
\end{figure}
We introduce a microscopic description of the fluid and the wall, as shown schematically on the left side of Fig.~\ref{fig:schematic illustration of the model}. The fluid consists of $N$ particles that are confined to an $L_x\times L_y \times L$ cuboid box.
We impose periodic boundary conditions along the $x$ and $y$ axes and introduce two walls so as to confine particles in the $z$ direction. We represent the two walls as collections of material points placed near planes $z=0$ and $z=L$, which are referred to as walls $A$ and $B$, respectively. Each wall consists of $N_A$ and $N_B$ material points.

The position and momentum of the $i$th particle of the fluid are denoted by $(\bm{r}_i,\bm{p}_i), \ (i=1,2,\cdots,N)$, and their collection is denoted by $\Gamma=(\bm{r}_1,\bm{p}_1,\cdots,\bm{r}_N,\bm{p}_N)$. Similarly, the position of the $i$th material point of wall $\alpha$ ($\alpha = A, B$) is denoted by $\bm{q}_i^{\alpha}, \ (i=1,2,\cdots,N_{\alpha})$ and their collection is denoted by $\Gamma^{\alpha} = (\bm{q}_1^{\alpha},\cdots,\bm{q}_{N_{\alpha}}^{\alpha})$.

The Hamiltonian of the system is given by
\begin{eqnarray}
& & \hspace{-0.5cm} H(\Gamma;\Gamma^A,\Gamma^B) = \sum_{i=1}^N \frac{\bm{p}_i^2}{2m} + \sum_{i<j} U_{FF}(|\bm{r}_i-\bm{r}_j|) \nonumber \\
+ & & \sum_{i=1}^{N} \sum_{j=1}^{N_A} U_{AF}(|\bm{r}_i-\bm{q}^A_j|) + \sum_{i=1}^{N} \sum_{j=1}^{N_B} U_{BF}(|\bm{r}_i-\bm{q}^B_j|).
\end{eqnarray}
$U_{\rm FF}(r)$ describes the interaction potential between the particles of the fluid and $U_{\rm \alpha F}(r)$ describes that between the material point of wall $\alpha$ and the fluid particle. In this paper, all the interaction potentials are given by the \modHN{modified} Lennard-Jones potential with a cut-off length $r_c$:
\begin{eqnarray}
U_{\alpha F}(r) = 4 \epsilon_{\alpha F} \Big\{\Big(\frac{\sigma_{\alpha F}}{r} \Big)^{12} - c_{\rm \alpha F}\Big(\frac{\sigma_{\alpha F}}{r} \Big)^{6} + C_{\alpha F}^{\rm (2)} r^2 + C_{\alpha F}^{\rm (0)} \Big\} \nonumber \\
\end{eqnarray}
for $r<r_c$, and $U_{\alpha F}(r) = 0$ otherwise, where $\alpha = F, A, B$. $C_{\alpha F}^{\rm (0)}$ and $C_{\alpha F}^{\rm (2)}$ are determined by the conditions $U_{\alpha F}(r_c) = 0$ and $U'_{\alpha F}(r_c) = 0$. \modHN{It should be noted that the potential in this form was frequently used in the molecular dynamics studies concerning the boundary condition~\cite{lauga2007microfluidics,cao2009molecular}.} For simplicity, we hereafter omit the subscript $FF$ in the parameters $(\sigma_{FF},c_{FF},\epsilon_{FF})$. Then, the time evolution of the system is described by Newton's equation:
\begin{eqnarray}
& & m \frac{d^2 \bm{r}_i}{dt^2}= - \sum_{j\neq i}\frac{\partial U_{FF}(|\bm{r}_i-\bm{r}_j|)}{\partial \bm{r}_i} \nonumber \\[3pt]
& & - \sum_{j=1}^{N_A}\frac{\partial U_{AF}(|\bm{r}_i-\bm{q}^A_j|)}{\partial \bm{r}_i} - \sum_{j=1}^{N_B}\frac{\partial U_{BF}(|\bm{r}_i-\bm{q}^B_j|)}{\partial \bm{r}_i}.
\label{eq:equation of motion: Hamilton}
\end{eqnarray}

The microscopic structure of wall $\alpha$ is given by the positions of the material points, $\Gamma^{\alpha}$, and the parameters in the interaction potential, $(\epsilon_{\alpha F}, \sigma_{\alpha F}, c_{\alpha F})$. In this paper, we study the atomically smooth walls given by the collection of material points that are fixed on the square lattice in the $z=0$ or $z=L$ planes. The lattice constant, which is denoted by $a_{\alpha}$, is given by
\begin{eqnarray}
\sigma_{\alpha F} = \frac{a_{\alpha}+\sigma}{2}
\end{eqnarray}
so that the lattice constant $a_{\alpha}$ is treated as the diameter of the particles constituting wall $\alpha$. Then, the microscopic structure of the wall is determined by other parameters $(a_{\alpha}, c_{\alpha F}, \epsilon_{\alpha F})$.

\subsection{Parameters}
\label{sec:parameters}
In numerical simulations, all the quantities are measured by unit $(m,\sigma,\epsilon)$. In particular, the time is measured by $\tau_{\rm micro}=\sqrt{m \sigma^2/\epsilon}$. For liquid argon, the length and time scales are $\sigma = 0.34{\rm nm}$ and $\tau_{\rm micro}=2.16\times 10^{-12}{\rm s}$~\cite{allen2017computer}.

The simulations are basically performed for $(L_x, L_y, L) =(50.0\sigma,50.0\sigma,20.0\sigma)$, $N=37500$, $c=1.0$, and $r_c=2.5\sigma$. 
\modHN{Although the wall potential is characterized by the three parameters $(a_{\alpha}, c_{\rm \alpha F}, \epsilon_{\rm \alpha F})$, in our study, we fix $(a_{\alpha},\epsilon_{\rm \alpha F})$ to $(0.6\sigma,0.9\epsilon)$ and examine three types of walls given by $c_{\rm \alpha F}=0.8,0.4,0.0$.} We refer to these walls as Wall I, II, and III, respectively.
The initial states are prepared using the Langevin thermostat with $k_B T/\epsilon = 2.0$. When solving Newton's equation (\ref{eq:equation of motion: Hamilton}), we use the second-order symplectic Euler method~\cite{yoshida1993recent} with a time step $dt=0.001\tau_{\rm micro}$. We note that for our parameters the fluid is in the supercritical state~\cite{baidakov2000effect,watanabe2012phase}.

\section{Non-equilibrium molecular dynamics simulation}
\label{subsec:Non-equilibrium molecular dynamics simulation}
Before proceeding the equilibrium molecular dynamics (EMD) simulation, we investigate the basic properties of the fluid and wall by using the non-equilibrium molecular dynamics (NEMD) simulation. \modHN{We simulate two types of flow, Couette flow and Poiseuille flow, and confirm the validity of the bulk constitutive equation and the partial slip boundary condition~\cite{barrat1999influence,barrat1999large}.}

\subsection{Method}
In Sec.~\ref{sec:Microscopic description of confined fluids}, we introduced the isolated Hamiltonian system. The non-equilibrium steady state is simulated by adding the Langevin thermostat and the external force. We study two types of flow. The first one is Poiseuille flow, which is realized by adding to all the particles the constant external force along the $x$-axis and the Langevin thermostat along the $y$-axis. That is, the particles obey the Langevin equation:
\begin{eqnarray}
m \frac{d^2 r_i^a}{dt^2} = - \frac{\partial H}{\partial r_i^a} + f^a + \delta_{ay}\big(- \zeta \frac{d r_i^a}{dt} + \xi_i^a(t)\big)
\end{eqnarray}
where $\delta_{ab}$ is Kronecker delta, $\bm{f}=(f,0,0)$ is the external force and $\bm{\xi}_i$ represents thermal noise satisfying
\begin{eqnarray}
\langle \xi_i^{\alpha}(t) \xi_j^{\beta}(t') \rangle = 2 \zeta k_{\rm B} T \delta_{ij} \delta^{\alpha \beta} \delta(t-t'),
\label{eq:langevin thermostat}
\end{eqnarray}
where $k_{\rm B}$ is the Boltzmann constant, $T$ the temperature of the thermostat, and $\zeta$ the friction coefficient. Here, we note that the Langevin thermostat controls the momentum only along the $y$-axis, which is orthogonal to the flow. In this simulation, we assume that walls $A$ and $B$ have the same microscopic structure.

The second one is Couette flow, which is simulated as follows. First, we replace wall $B$ from the collection of the material points to the following potential
\begin{eqnarray}
V_{B}(\bm{r}) &=& 4 \epsilon \Big\{\big(\frac{\sigma}{L-z}\big)^{12} - \big(\frac{\sigma}{L-z}\big)^{6} \nonumber \\[3pt]
& & \hspace{1cm} + C^{\rm (2)}_{FF} (L-z)^2 + C^{\rm (0)}_{FF}\Big\}
\end{eqnarray}
for $r<r_c$, and $V_B(\bm{r}) = 0$ otherwise. Then, the tangential momentum is not exchanged between the fluid and wall $B$ because the force acting on the fluid particle $i$ is calculated as $-\partial V_B(\bm{r}_i)/\partial r_i^a$.  Second, we apply the Langevin thermostat and the external force along the $x$-axis in the region near wall $B$ (referred to as region $\mathcal{R}_B$). Specifically, in region $\mathcal{R}_B$, the particles obey the Langevin equation
\begin{eqnarray}
m \frac{d^2 r_i^a}{dt^2} = - \frac{\partial H}{\partial r_i^a} + f^a - \zeta \frac{d r_i^a}{dt} + \xi_i^a(t)
\end{eqnarray}
where $\bm{f}=(f,0,0)$ and $\bm{\xi}_i$ is given by (\ref{eq:langevin thermostat}). Outside region $\mathcal{R}_B$, we use the same dynamics as the EMD simulation. We note that in the Couette flow the Langevin thermostat controls all components of momentum. In the steady state, the velocity of fluid in region $\mathcal{R}_B$ is approximately given by $\bm{v} = \bm{f}/\zeta$.

For both types of flow, we concentrate on the mass density field $\rho(\bm{r})$, the $x$-component of the velocity field $v^x(\bm{r})$ and the shear stress field $\sigma^{xz}(\bm{r})$ in the steady state. These quantities are calculated from the microscopic mass density field $\hat{\rho}(\bm{r};\Gamma)$, the momentum density field $\hat{\pi}^a(\bm{r};\Gamma)$ and the momentum current density field $\hat{J}^{ab}(\bm{r};\Gamma)$. These microscopic quantities are defined in terms of microscopic configuration $\Gamma$. The details are referred to Ref.~\cite{nakano2019microscopic}. Because we are interested in the $z$-dependence of the averaged local quantities, we perform spatial average in the slab with a bin width $\Delta z$ at the center $z$ and temporal average for a time interval $\Delta T_{\rm obs}$ in the steady state. For example, the averaged mass density at any $z$ is given by
\begin{eqnarray}
\rho(z) &=& \frac{1}{\Delta T_{\rm obs}} \int_{0}^{\Delta T_{\rm obs}} ~\hspace{-0.3cm} dt  \nonumber \\
& & \hspace{0.2cm}\frac{1}{L_xL_y} \int dxdy  \frac{1}{\Delta z} \int_{z-\Delta z/2}^{z+\Delta z/2} dz \hat{\rho}(\bm{r};\Gamma_t),
\label{eq:example of rho}
\end{eqnarray}
where the system is assumed to be in steady state at $t=0$. The remaining quantities are calculated in the same way~\cite{nakano2019microscopic}. We choose $\Delta T_{\rm obs}=10000\tau_{\rm micro}$. Furthermore, we take an ensemble average over 8 different initial states.

\subsection{Density profile and velocity profile}
\begin{figure*}[]
\begin{center}
\subfigure[]{
\includegraphics[width=8cm]{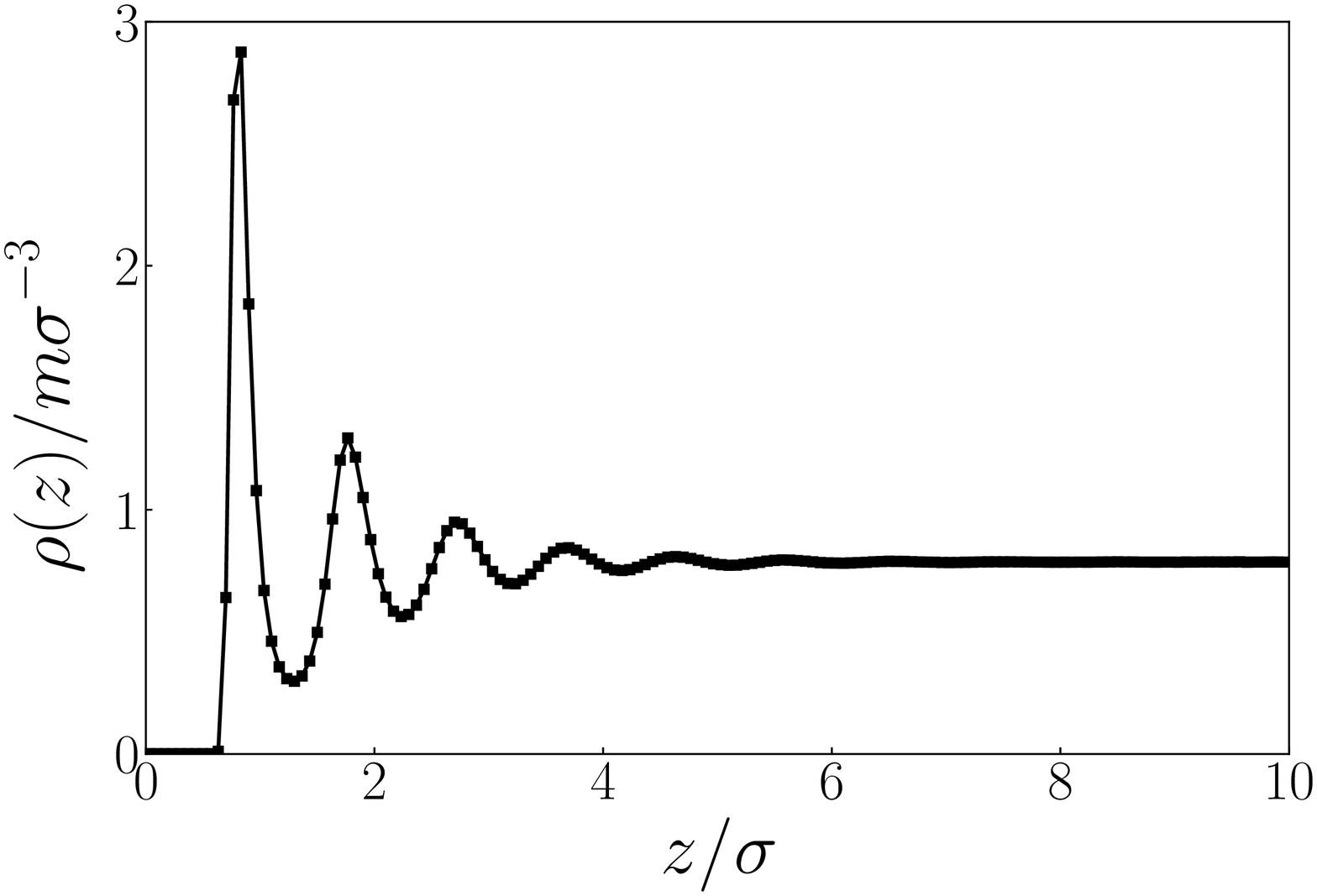}
\label{fig:density profile}}
\subfigure[]{
\includegraphics[width=8cm]{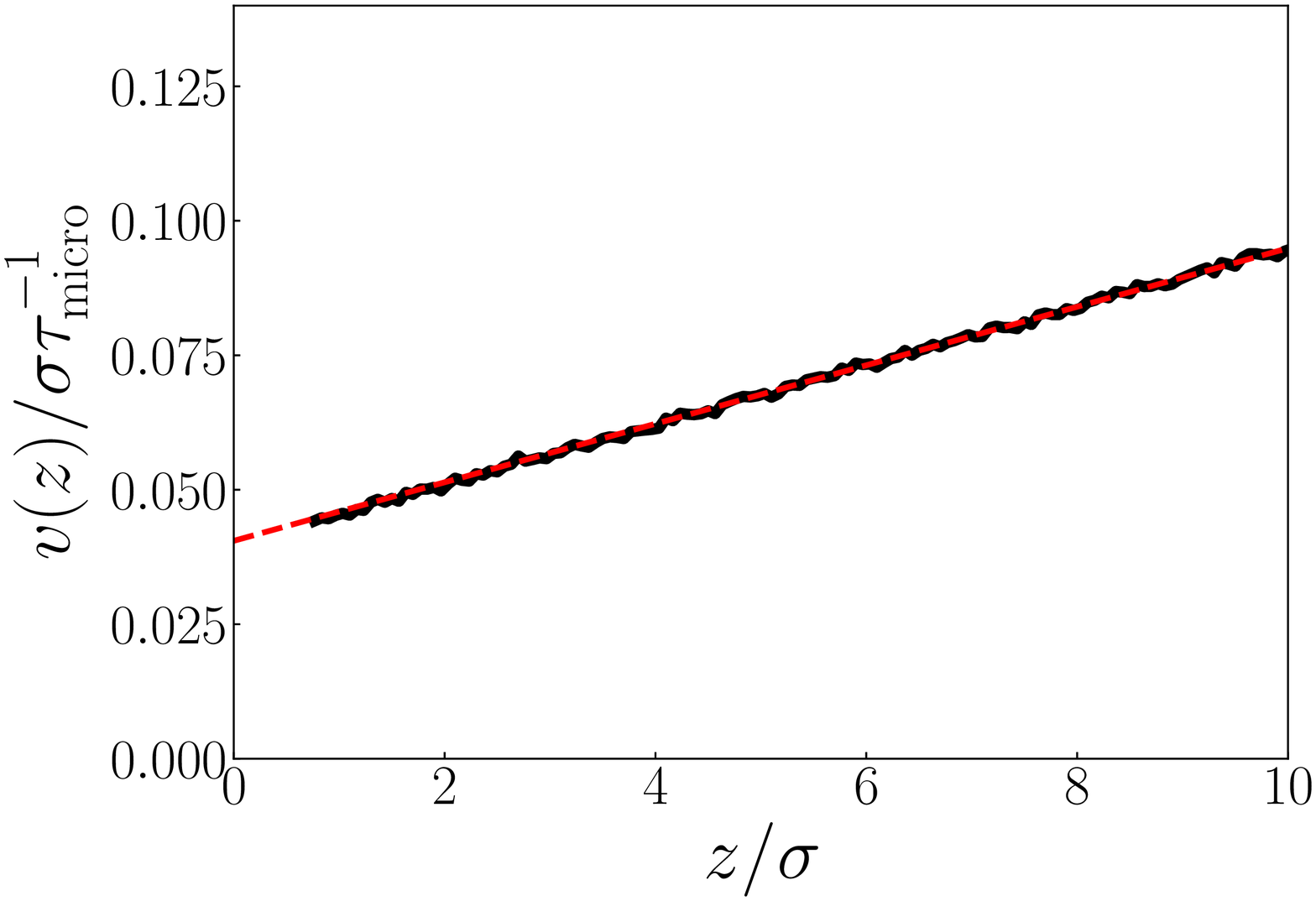}
\label{fig:velocity profile}}
\end{center}
\vspace{-0.5cm}
\caption{(Color online) (a) density profile and (b) velocity profile in the non-equilibrium steady state under Couette flow. The wall parameter is set to  \modHN{$c_{AF}=0.8$} (Wall I), and the external force $f=0.15\epsilon/\sigma$ is imposed.}
\label{fig:typical snapshot}
\end{figure*}

We present the density and velocity profiles in Fig.~\ref{fig:typical snapshot}, which are obtained for the Couette flow with the external force $f=0.15\epsilon/\sigma$. The wall parameter is set to  \modHN{$c_{AF}=0.8$} (Wall I). These figures show typical features that are observed for the atomically smooth wall. First, from Fig.~\ref{fig:density profile}, we find that the density profile is not uniform but oscillates near the wall. Such an oscillation is generally observed and the amplitude is known to depend on the strength of attraction of the walls~\cite{thompson1990shear,barrat1999large,voronov2008review}. The density profile becomes uniform away from the walls. For our models, the oscillation of density is not observed for $z>6.0\sigma$. Thus, we refer to this region as bulk. Next, we focus on the velocity profile in Fig.~\ref{fig:velocity profile}, which is obtained for the same setup as the density profile. The black solid curve and the red dashed line, respectively, represent the simulation data and the fitting result. The detailed method of the fitting will be explained in the next subsection. The fitting result shows that the velocity profile is not singular near the wall; specifically, it is fitted by the same linear function both in the bulk region and near the wall.

\subsection{Fitting method of velocity profile}
\label{sec:fitting method of velocity profile}
The velocity profile is fitted to the solution of Navier--Stokes equation with the partial slip boundary condition. Here, there are three fitting parameters. The first one is the viscosity $\eta$, which involves the bulk constitutive equation:
\begin{eqnarray}
\sigma^{xz} = \eta \frac{dv^x}{dz}.
\label{eq:bulk constitutive equation}
\end{eqnarray}
The other two parameters are contained in the partial slip boundary condition. As explained in Introduction, the partial slip boundary condition is characterized by the slip length. However, more precisely, one more parameter is required to describe the partial slip boundary condition. It is the hydrodynamic position of the fluid-wall interface, which is defined as the position where the boundary condition is imposed. Explicitly, the partial slip boundary condition is written as
\begin{eqnarray}
v^x(z_w) = b(z_w) \frac{\partial v^x(z)}{\partial z} \Big|_{z=z_w} 
\label{eq:partial slip boundary condition:exact form}
\end{eqnarray}
where $z_w$ is the hydrodynamic wall position and $b(z_w)$ is the slip length. It should be noted that the slip length depends on the hydrodynamic wall position $z_w$~\cite{bocquet1994hydrodynamic}. Therefore, we express this dependence as $b(z_w)$. 

Furthermore, we point out that the slip length depends on the flow geometry. This fact is immediately understood from the following argument as discussed in Ref.~\cite{bocquet1994hydrodynamic}. For the Couette flow, the velocity field is given by
\begin{eqnarray}
v^x(z) = \dot{\gamma}\big(z+b^c(z_w)\modHN{-z_w}\big),
\label{eq:velocity field: Couette flow}
\end{eqnarray}
where $\dot{\gamma}$ is the shear rate and $b^c(z_w)$ represents the slip length of the Couette flow. From (\ref{eq:partial slip boundary condition:exact form}) and (\ref{eq:velocity field: Couette flow}), we find that $b^c(z_w)$ depends on $z_w$ as
\begin{eqnarray}
b^c(z_w) = b^c(0)+z_w.
\label{eq:zw dependency of slip length: couette flow}
\end{eqnarray}
Next, for the Poiseuille flow, the velocity field is calculated as
\begin{eqnarray}
v^x(z) &=& \frac{f}{2\eta}z(L-z) \nonumber \\[3pt]
&\modHN{-}& \frac{f}{2\eta}\big(z_w(L-z_w)-b^p(z_w)(L-2z_w)\big),
\label{eq:velocity field: Poiseuille flow}
\end{eqnarray}
where $f$ is the external force and $b^p(z_w)$ represents the slip length of the Poiseuille flow. Accordingly, the slip length satisfies
\begin{eqnarray}
b^p(z_w) = b^p(0)+z_w+\frac{z_w(z_w+2b^p(0))}{L-2z_w}
\label{eq:zw dependency of slip length: poiseuille flow}
\end{eqnarray}
where $z_w<L/2$. Clearly, there is the difference between the $z_w$-dependence of (\ref{eq:zw dependency of slip length: couette flow}) and (\ref{eq:zw dependency of slip length: poiseuille flow}). Therefore, the slip lengths $b^c(z_w)$ and $b^p(z_w)$ do not generally coincide.

Now we impose that the proper definition of the slip length and the hydrodynamic wall position does not depend on the flow geometry. When we focus on the two types of geometry, this condition can be satisfied. Specifically, we choose the hydrodynamic wall position by solving the equation
\begin{eqnarray}
b^c(z_w^{\ast}) = b^p(z_w^{\ast}) = b^{\ast}.
\label{eq:well-defined hydrodynamic position}
\end{eqnarray}
The solution of this equation is calculated as
\begin{eqnarray}
z_w^{\ast} &=& - b^c(0) + \sqrt{(b^c(0))^2+L(b^c(0)-b^p(0))}
\label{eq:def of well-defined hydrodynamic position},\\
b^{\ast} &=&  \sqrt{(b^c(0))^2+L(b^c(0)-b^p(0))},
\label{eq:def of well-defined slip length}
\end{eqnarray}
which provides the operational definition of the slip length and the hydrodynamic wall position.

In summary, the fitting is performed in the following steps. First, we observe the velocity field $v^x(z)$ and the shear stress field $\sigma^{xz}(z)$. Second, we fit the velocity field $v^x(z)$ in the bulk region to (\ref{eq:velocity field: Couette flow}) and (\ref{eq:velocity field: Poiseuille flow}) with $z_w=0$. The slip length $b(0)$ is obtained by substituting the fitting result into (\ref{eq:partial slip boundary condition:exact form}). $(z_w^{\ast},b^{\ast})$ is calculated from (\ref{eq:def of well-defined hydrodynamic position}) and (\ref{eq:def of well-defined slip length}). The viscosity $\eta$ is calculated by using the bulk constitutive equation (\ref{eq:bulk constitutive equation}). \modHN{This method was used in the literatures~\cite{barrat1999large,barrat1999influence,pastorino2006static,muller2007cyclic,muller2008flow,pastorino2009coarse}.}

\subsection{Fitting result}
\label{sec:fitting result}
\begin{figure*}[]
\begin{center}
\subfigure[]{
\includegraphics[width=8cm]{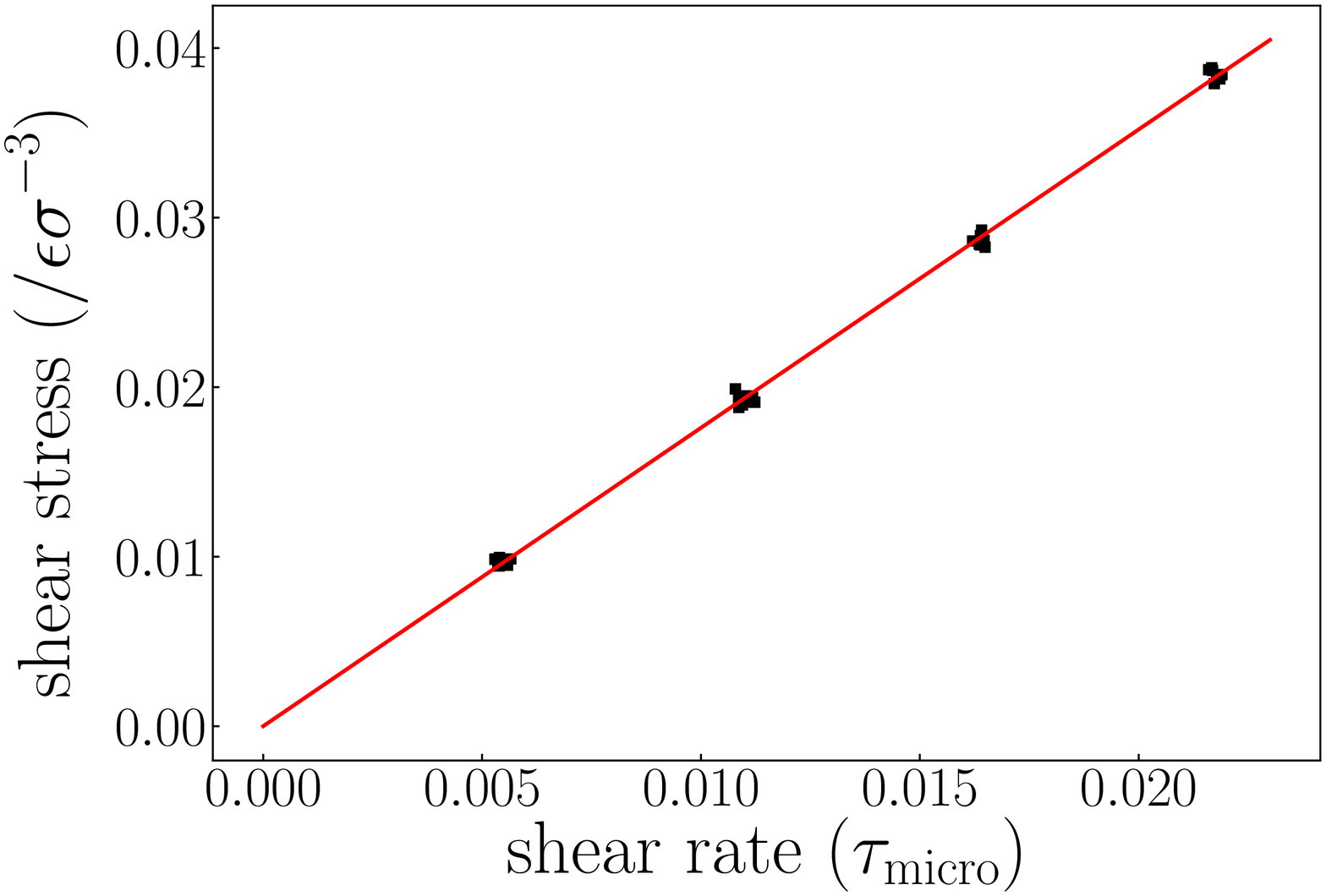}
\label{fig:bulk constitutive equation}}
\subfigure[]{
\includegraphics[width=8cm]{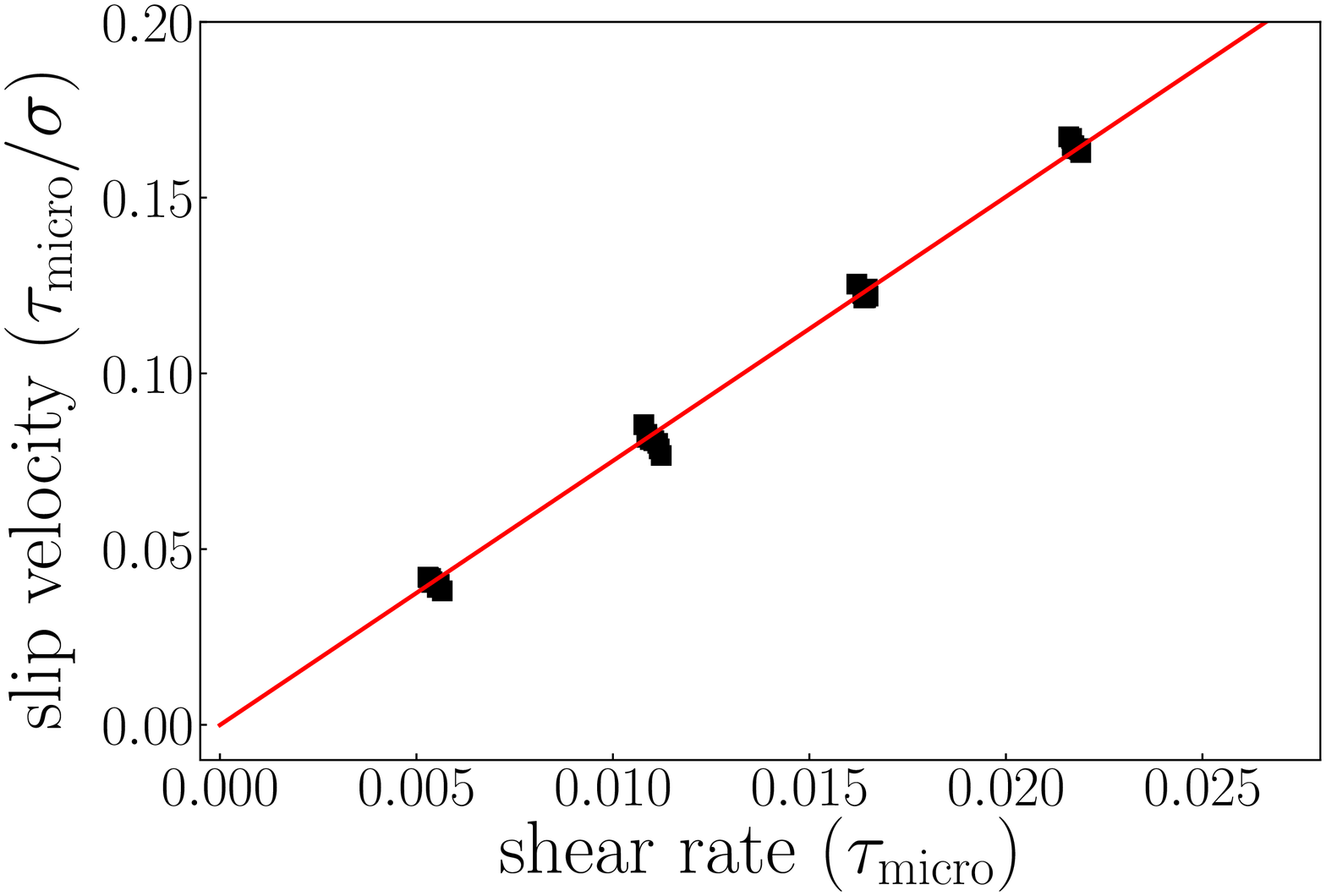}
\label{fig:boundary constitutive equation}}
\end{center}
\vspace{-0.7cm}
\caption{(Color online) (a) shear stress $\sigma^{xz}$ as a function of shear rate $\dot{\gamma}$. (b) slip velocity $v^x(0)$ as a function of shear rate $\dot{\gamma}$. The observation is performed for Couette flow with Wall I \modHN{$(c_{AF}=0.8)$}. The red line in (a) is $\sigma^{xz} = 1.76\dot{\gamma}$. The red line in (b) is $v^x(0) = 8.08\dot{\gamma}$.}
\end{figure*}

First, we validate the bulk constitutive equation (\ref{eq:bulk constitutive equation}). In Fig.~\ref{fig:bulk constitutive equation}, the shear stress is plotted as the function of the shear rate for the Couette flow. The wall is set to Wall I (same as Fig.\ref{fig:typical snapshot}). This plot is well fitted by (\ref{eq:bulk constitutive equation}) with $\eta=1.76$, and then we confirm the validity of (\ref{eq:bulk constitutive equation}). Here, we note that the bulk constitutive equation holds even near the walls and the viscosity $\eta$ hardly depends on the $z$-coordinate, because the velocity profile does not exhibit the singular behavior close to the wall (see the left side of Fig.~\ref{fig:typical snapshot}).

Next, we validate the partial slip boundary condition (\ref{eq:partial slip boundary condition:exact form}). In Fig.~\ref{fig:boundary constitutive equation}, the slip velocity is plotted as the function of the shear rate for the same setup as Fig.~\ref{fig:bulk constitutive equation}. Here, the slip velocity and the shear rate are measured from the velocity field extrapolated to $z=0$, and then the slope gives the slip length $b^c(0)$. From the fact that this plot can be fitted with the linear function, we confirm the validity of the partial slip boundary condition (\ref{eq:partial slip boundary condition:exact form}).

These two results also hold for the Poiseuille flow with all types of wall that we examined. Accordingly, we conclude that our model obeys the Navier--Stokes equation with the partial slip boundary condition. We summarize the values of parameters obtained from the NEMD simulation in Table~\ref{tab:density and viscosity in bulk region}. The error represents the standard deviation for different 8 initial states. We distinguish the quantities obtained from the NEMD simulation with subscript neq. $b_{\rm neq}$ and $z_{\rm neq}$ are well-defined quantities, which are calculated from (\ref{eq:well-defined hydrodynamic position}).

Below, we will argue an equilibrium measurement method of the slip length by using the same walls as used here. Before proceeding to the equilibrium measurement, we comment on the values of $b_{\rm neq}$ and $z_{\rm neq}$ that we obtained. The slip length $b_{\rm neq}$ takes the wide range of values. For Walls I and II, the slip length $b_{\rm neq}$ is comparable to the system size $L$. Furthermore, for Wall III, the slip length $b_{\rm neq}$ is about 6 times larger than the system size $L$. We note that such phenomena have been observed due to the recent development of nanofluidics~\cite{falk2010molecular,kannam2013fast,secchi2016massive}. 
Concerning the hydrodynamic wall position $z_{\rm neq}$, some group reported that it is located near the first peak of the density profile (see Fig.~\ref{fig:density profile})~\cite{barrat1999large,chen2015determining,PhysRevFluids.4.114201}. In our simulation, for Wall I, the first peak is observed at $z_{\rm peak} \sim 0.83\sigma$ which is smaller than $z_{\rm neq}$. It should be noted that this difference is much smaller than the slip length.

\begin{table*}[ht]
\begin{center}
\begin{tabularx}{180mm}{C||CCC}\hline
Wall & $\eta_{\rm neq}/\sqrt{m\epsilon}\sigma^{-2}$ & $b_{\rm neq}/\sigma$ & $z_{\rm neq}/\sigma$ \\ \hline\hline
I \ \  \modHN{$(c_{AF}=0.8)$} & $1.76\pm0.01$ & $8.08\pm0.14$ & $0.56\pm0.09$ \\
II \ \  \modHN{$(c_{AF}=0.4)$} &$1.79\pm0.02$ & $30.8\pm0.3$ & $0.43\pm0.11$\\
III \ \  \modHN{$(c_{AF}=0.0)$} & $1.81\pm0.03$ & $121.8\pm1.7$ & $1.1\pm0.2$ \\\hline\hline
\end{tabularx}
\caption{Viscosity $\eta_{\rm neq}$, slip length $b_{\rm neq}$ and hydrodynamic wall position $z_{\rm neq}$, which are obtained in the NEMD simulation. The error represents standard deviation for 8 different initial states.}
\label{tab:density and viscosity in bulk region}
\end{center}
\end{table*}

\section{Review of BB's relation}
\label{sec:Review of BB's formula}
In this section, we review the calculation method of BB's relation and sort out the problems with this method.

\subsection{Method}
BB's relation is calculated from the EMD simulation. Then, we numerically solve Newton's equation (\ref{eq:equation of motion: Hamilton}). The microscopic expression of the force acting on wall $\alpha$ is given by
\begin{eqnarray}
\hat{\bm{F}}_{\alpha}(\Gamma) = \sum_{j=1}^{N_A}\frac{\partial U_{\alpha F}(|\bm{r}_i-\bm{q}^{\alpha}_j|)}{\partial \bm{r}_i}.
\end{eqnarray}
The force autocorrelation function is calculated by performing a long-time average over the time interval $\Delta T_{\rm obs}$:
\begin{eqnarray}
\hspace{-0.5cm}\langle F^a_{\alpha}(t) F^a_{\alpha}(0) \rangle_{\rm eq} = \frac{1}{\Delta T_{\rm obs}} \int_0^{\Delta T_{\rm obs}} \hspace{-0.3cm}ds\hspace{0.01cm} \hat{F}^a_{\alpha}(\Gamma_{t+s})\hat{F}^a_{\alpha}(\Gamma_s),
\label{eq:calculation method of force autocorrelation function}
\end{eqnarray}
where $a=x, y, z$, and the system is assumed to be in equilibrium at $t=0$. We choose $\Delta T_{\rm obs}=10000\tau_{\rm micro}$, which is the same as the NEMD simulation. We also take an ensemble average over 24 different initial states. We introduce the Green--Kubo integral of the force autocorrelation function:
\begin{eqnarray}
\gamma^{aa}_{{\rm MD},\alpha \alpha}(t) = \frac{1}{k_B T L_xL_y}\int_0^t ds \langle F^a_{\alpha}(s) F^a_{\alpha}(0) \rangle_{\rm eq}.
\label{eq:gamma(t) in MD}
\end{eqnarray}
Here, the subscript MD represents that $\gamma^{aa}_{{\rm MD},\alpha \alpha}(t)$ is calculated from the EMD simulation. For simplicity, we focus on the $x$-component of the force $\hat{F}^x_{\alpha}$ and omit the superscript $x$ in $\hat{F}^x_{\alpha}$ and $\gamma^{xx}_{{\rm MD},\alpha \alpha}(t)$. We assume that walls $A$ and $B$ have the same microscopic structure so that $(a_{A}, c_{AF}, \epsilon_{AF}) = (a_{B}, c_{BF}, \epsilon_{BF})$. In this case, it holds that
\begin{eqnarray}
\langle F_A(t)F_A(0) \rangle_{\rm eq} = \langle F_B(t)F_B(0) \rangle_{\rm eq}
\end{eqnarray}
for any $t$. We thus study only the behavior near wall $A$ and drop the subscript $A$ from $\hat{F}_{A}(\Gamma)$ and $\gamma_{{\rm MD}, AA}(t)$.

As explained in Introduction, Bocquet and Barrat proposed that the friction coefficient $\lambda$ is calculated as
\begin{eqnarray}
\lambda = \gamma_{\rm MD}(\tau_0)
\label{eq:BB's relation revisted1}
\end{eqnarray}
where $\tau_0$ is defined as the first zero of $\langle F(t)F(0) \rangle_{\rm eq}$~\footnote{
This method was originally developed in the context of Stokes' law; J. Brey and J. G. Ord\'o\~nez, J. Chem. Phys. 76, 3260 (1982); P. Espa\~nol and I. Z\'u\~niga, J. Chem. Phys. 98, 574 (1993)
}. As mentioned in Sec.~\ref{sec:fitting method of velocity profile}, the value of the slip length depends on the position where the slip velocity is defined. Because the friction coefficient is defined by (\ref{eq:def of the friction coefficient}), we notice that the same problem occurs for the friction coefficient. Then, in order to calculate the friction coefficient from the relation (\ref{eq:BB's relation revisted1}), we must determine the position of the fluid-wall interface.

In this paper, we assume that the friction coefficient calculated from the relation (\ref{eq:BB's relation revisted1}) yields the flow-geometry-independent value. Specifically, the position of the fluid-wall interface is given by the hydrodynamic wall position $z_{\rm neq}$. This assumption is justified if the relation (\ref{eq:BB's relation revisted1}) is connected with the Green--Kubo formula, because we can derive that the Green--Kubo formula provides the flow-geometry-independent friction coefficient~\cite{nakano2019statistical}. However, because the connection between the relation (\ref{eq:BB's relation revisted1}) and the Green--Kubo formula is unclear so far, as mentioned in Introduction, this choice is just an assumption (see also Sec.~\ref{subsec:Problem and strategy}).

\subsection{BB's relation}
\begin{figure*}[ht]
\begin{center}
\hspace{-0.2cm}
\subfigure[]{
\includegraphics[width=8cm]{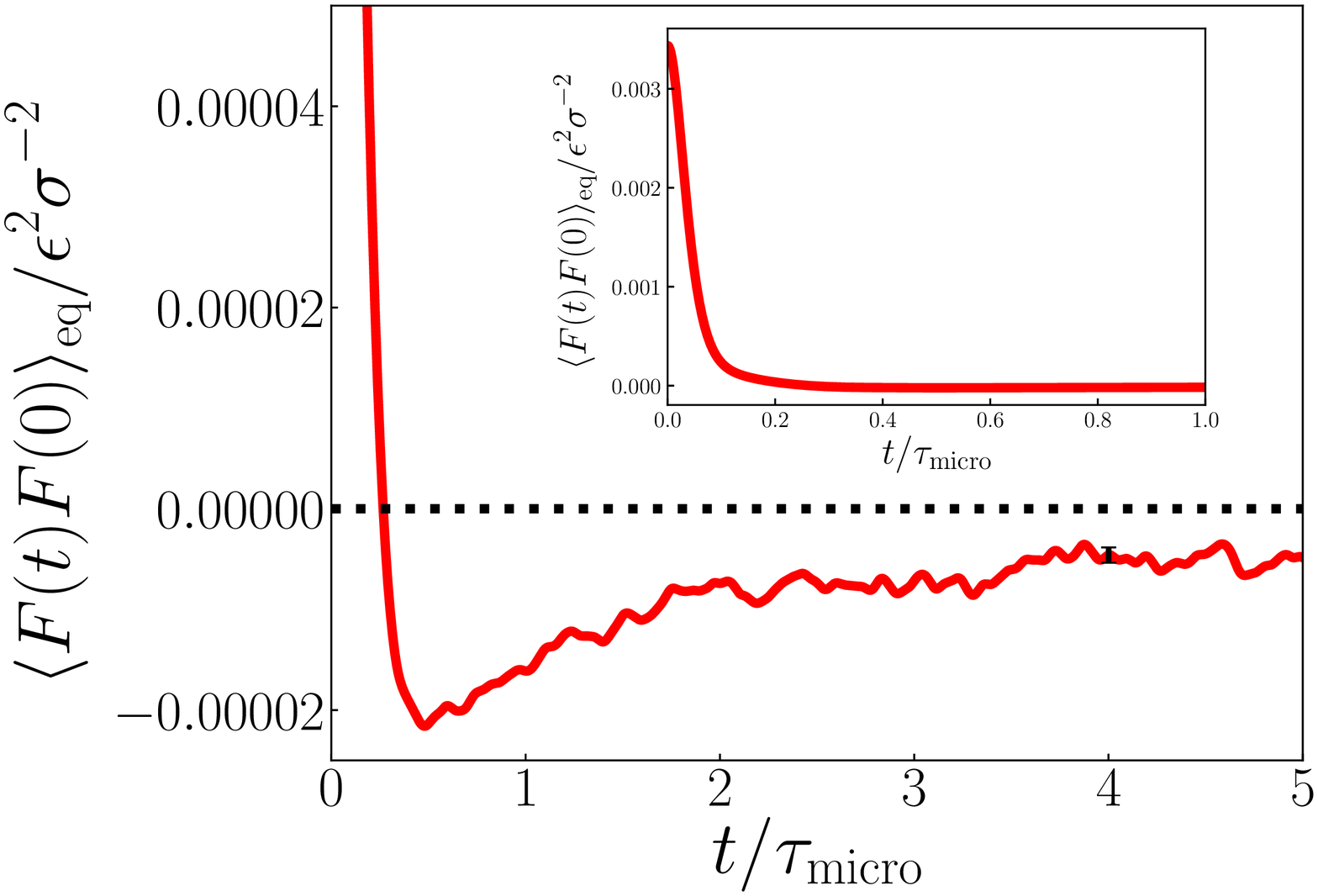}
\label{fig:typical simulation result: aw0.6cw0.8: FAF}}\hspace{0.2cm}
\subfigure[]{
\includegraphics[width=8cm]{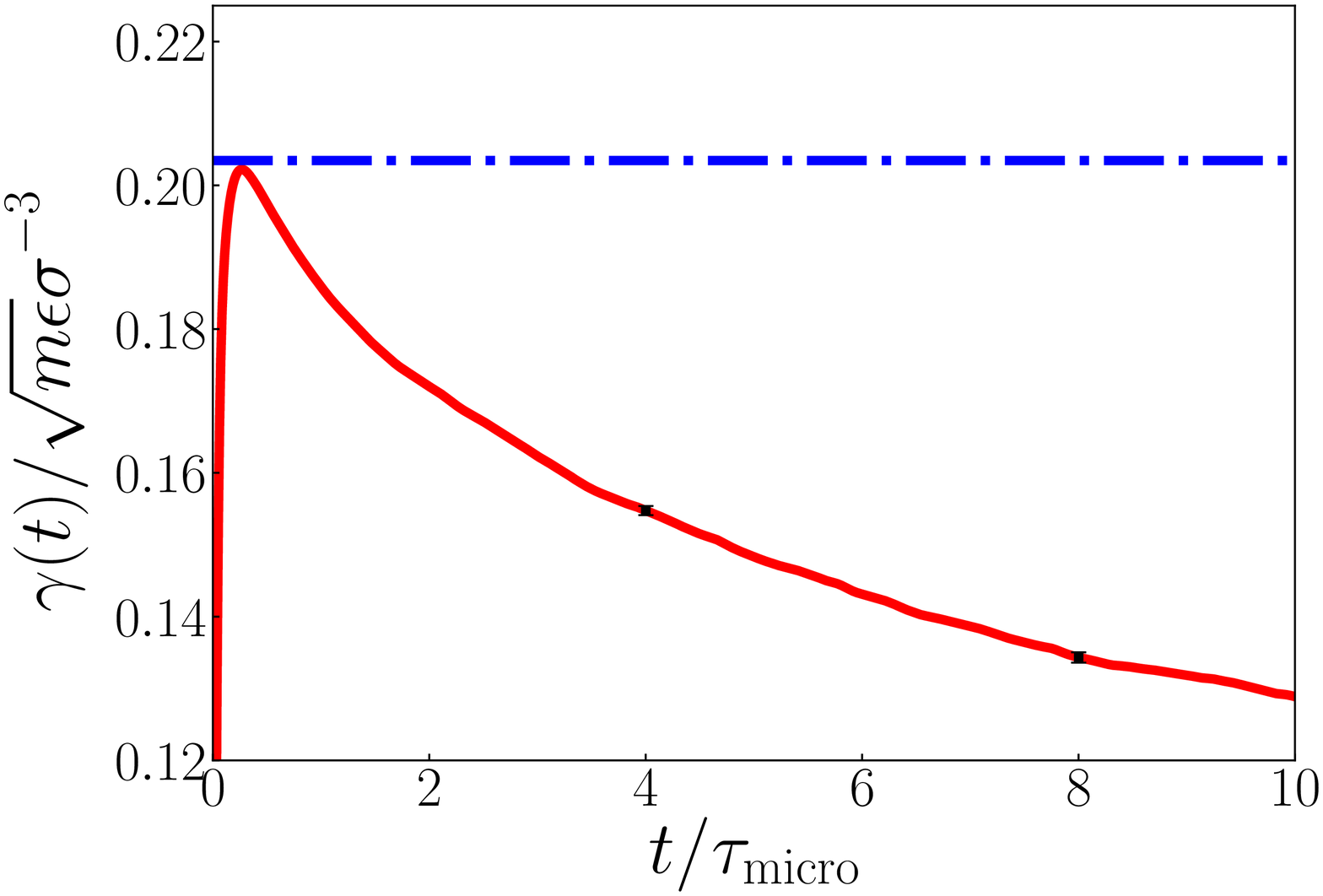}
\label{fig:typical simulation result: aw0.6cw0.8: gamma}} 
\end{center}
\vspace{-0.7cm}
\caption{(Color online) (a) $\langle F(t)F(0) \rangle_{\rm eq}$ and (b) $\gamma_{\rm MD}(t)$ for Wall I, which are calculated in the molecular dynamics simulation.  The inset in (a) is an expansion of time region $[0,\tau_{\rm micro}]$. The blue dash-dot line in (b) is the value of the first peak of $\gamma_{\rm MD}(t)$.}
\label{fig:typical simulation result: aw0.6cw0.8}
\end{figure*}

In Fig.~\ref{fig:typical simulation result: aw0.6cw0.8}, we present $\langle F(t)F(0) \rangle_{\rm eq}$ and $\gamma_{\rm MD}(t)$ for Wall I. For this wall, the first zero of $\langle F(t)F(0) \rangle_{\rm eq}$ is about $\tau_0\sim 0.3\tau_{\rm micro}$ (see Fig.~\ref{fig:typical simulation result: aw0.6cw0.8: FAF}), which is comparable to the characteristic time scale of molecular motion. For example, when two particles are bound by the attractive interaction $U_{FF}(r)$, the period of the vibration is calculated as $0.83\tau_{\rm micro}$. The relation (\ref{eq:BB's relation revisted1}) implies that the friction coefficient is calculated from information over the molecular time scale.

Although the relation (\ref{eq:BB's relation revisted1}) is immediately calculated from the EMD simulation, we require the additional assumption in order to connect the friction coefficient $\lambda$ with the slip length $b$. It is that the bulk constitutive equation (\ref{eq:bulk constitutive equation}) holds adjacent to the wall. In Sec.~\ref{sec:fitting result}, we showed the validity of this assumption by using the NEMD simulation. Then, by using (\ref{eq:bulk constitutive equation}), the slip length is expressed as
\begin{eqnarray}
b _{\rm peak}= \frac{\eta}{\gamma_{\rm MD}(\tau_0)} = \frac{\eta}{\lambda}.
\label{eq:BB's relation revisted2}
\end{eqnarray}
As mentioned in Introduction, we refer to this relation as BB's relation. Because the first zero of $\langle F(t)F(0) \rangle_{\rm eq}$ corresponds to the first peak of $\gamma_{\rm MD}(t)$ (see Fig.~\ref{fig:typical simulation result: aw0.6cw0.8: gamma}), we denote by $b_{\rm peak}$ the slip length related to BB's relation.

In Table~\ref{tab:result of BB's relation}, we give the value of $\lambda$ and $b_{\rm peak}$ for the three types of walls. In order to obtain $b_{\rm peak}$ from (\ref{eq:BB's relation revisted2}), we use the viscosity $\eta_{\rm neq}$ measured in the NEMD simulation (see Table~\ref{tab:density and viscosity in bulk region}). The error of $\lambda$ is the standard deviation for different 24 initial states. The error of $b_{\rm peak}$ is estimated from the propagation of error of BB's relation (\ref{eq:BB's relation revisted2}). The error of $\lambda$ is extremely small because the relation (\ref{eq:BB's relation revisted1}) is calculated from the information at the molecular time scale. As a result, the error for $b_{\rm peak}$ mainly comes from that of the viscosity $\eta_{\rm neq}$. By comparing Tables~\ref{tab:density and viscosity in bulk region} and \ref{tab:result of BB's relation}, we find that $b_{\rm peak}$ and $b_{\rm neq}$ are generally consistent over a wide region of the slip length, with deviations within $15\%$. This result is consistent with previous studies~\cite{bocquet1994hydrodynamic,falk2010molecular,tocci2014friction,wei2014breakdown,liang2015slip,ramos2016hydrodynamic}. If we allow for these deviations, we can obtain an estimation of the slip length via BB's relation.
\begin{table}[h]
\begin{center}
\begin{tabularx}{85mm}{c||CC}\hline
Wall  & $\lambda$ & $b_{\rm peak}$ \\ \hline\hline
I \ \   \modHN{$(c_{AF}=0.8)$} & $0.202\pm0.00$ & $8.71\pm0.05$ \\
II \ \   \modHN{$(c_{AF}=0.4)$} & $0.060\pm0.00$ & $29.8\pm0.3$ \\
III \ \   \modHN{$(c_{AF}=0.0)$} & $0.0168\pm0.00$ & $107.7\pm1.8$   \\\hline\hline
\end{tabularx}
\caption{friction coefficient $\lambda$ and slip length $b_{\rm peak}$ obtained from the relations (\ref{eq:BB's relation revisted1}) and (\ref{eq:BB's relation revisted2}). $b_{\rm peak}$ is calculated using the viscosity $\eta_{\rm neq}$. The error for $\lambda$ is the standard deviation for 24 different initial states. The error of $b_{\rm peak}$ is estimated from the propagation of error of BB's relation (\ref{eq:BB's relation revisted2}).} 
\label{tab:result of BB's relation}
\end{center}
\end{table}

\subsection{Problem and strategy}
\label{subsec:Problem and strategy}
The essential problem with the relation (\ref{eq:BB's relation revisted1}) is the lack of the clear theoretical understanding. A number of studies attempted to connect the relation (\ref{eq:BB's relation revisted1}) with the Green--Kubo formula. However, as mentioned in Introduction, because two types of limit are coupled in the Green--Kubo formula, it is difficult to validate the relation between the relation (\ref{eq:BB's relation revisted1}) and the Green--Kubo formula. Associated with this theoretical problem, two questions arise when we calculate the slip length $b$ from BB's relation (\ref{eq:BB's relation revisted2}): (i) why does $\tau_0$ provide the slip length?, (ii) where is the hydrodynamic wall position located?

The goal of this paper is to answer the question (i). Our crucial idea to tackle this problem is to examine how $\gamma_{\rm MD}(t)$ decays from $t=\tau_0$ (see Fig.~\ref{fig:typical simulation result: aw0.6cw0.8: gamma}) by introducing LFH as a phenomenological model of the confined fluid. We first demonstrate that LFH describes the fluctuations even at the molecular scale with high accuracy, although LFH was originally developed to describe the macroscopic fluctuations~\cite{landau1959course}. Then, by combining the EMD simulation result with the LFH solution, we propose a new theoretical interpretation of BB's relation (\ref{eq:BB's relation revisted2}), which answers the question (i). We also propose a new equilibrium measurement method for the slip length (or equivalently the friction coefficient). Although our method is expected to solve the question (ii), it is not attained in this paper mainly because of the measurement accuracy. We will discuss this point in Sec.~\ref{subsec:early-time behavior of gamma} and Appendix~\ref{subsec:discussion}.

\section{phenomenological description of confined fluid}
\label{sec:universal model for force autocorrelation function}
If the length scale of the spatial variation of the velocity field is much larger than molecular scales, the molecular motion is smoothed out and the fluid motion is described as a continuum. From this consideration, LFH is introduced to describe the thermal fluctuation and the fluid flow. In this section, we explain some exact results derived from LFH~\cite{nakano2019statistical}. In the next section, we will compare the EMD simulation results with the LFH predictions.

\subsection{Linearized fluctuating hydrodynamics (LFH)}
\label{subsec:Linearized fluctuating hydrodynamics}
We consider the same geometry of the system as that in Sec.~\ref{subsec:model}. The fluid is assumed to be incompressible. Accordingly, the time evolution is described by the incompressible Navier--Stokes equation with stochastic fluxes,
\begin{eqnarray}
\rho \frac{\partial \tilde{v}^a}{\partial t} + \frac{\partial \tilde{J}^{ab}}{\partial r^b} = 0,
\label{eq:Navier-Stokes equation subjected to a Gaussian random stress tensor}
\end{eqnarray}
where twice repeated indices are assumed to be summed over. The momentum flux tensor $\tilde{J}^{ab}(\bm{r},t)$ is given by
\begin{eqnarray}
\tilde{J}^{ab} =  \tilde{p} \delta_{ab} - \eta \Big(\frac{\partial \tilde{v}^a}{\partial r^b}  + \frac{\partial \tilde{v}^b}{\partial r^a}\Big) + \tilde{s}^{ab},
\label{eq:momentum flux tensor}
\end{eqnarray}
where $\tilde{s}^{ab}(\bm{r},t)$ is the Gaussian random stress tensor satisfying
\begin{eqnarray}
& & \big\langle \tilde{s}^{ab}(\bm{r},t) \tilde{s}^{cd}(\bm{r}',t') \big\rangle_{\rm eq} \nonumber \\[3pt] &=& 2 k_B T \eta \Big[\delta_{ac}\delta_{bd} + \delta_{ad}\delta_{bc} - \frac{2}{3} \delta_{ab} \delta_{cd} \Big] \delta^3(\bm{r}-\bm{r}') \delta(t-t').\nonumber \\
\label{eq:random stress tensor}
\end{eqnarray}
Here, the nonlinear effect induced by the advection term is ignored.

$\rho$ represents the fluid density, which corresponds to (\ref{eq:example of rho}) in the MD simulation. In LFH, $\rho$ is assumed to be spatially constant because of the incompressibility condition. This is not consistent with the observation in the MD simulation because the density profile close to the wall is not uniform but oscillates (see Fig.~\ref{fig:density profile}). This oscillation exists only close to the wall; if the system size is sufficiently large, the density profile is uniform in almost all of the region. Then, we interpret that $\rho$ in LFH is given by the density in such region. $\tilde{p}$ is the pressure, which is used to enforce the incompressibility condition,
\begin{eqnarray}
\frac{\partial \tilde{v}^a(\bm{r},t)}{\partial r^a} = 0.
\label{eq:divergence free condition}
\end{eqnarray}

We impose periodic boundary conditions along the $x$- and $y$-axes. The boundary conditions at walls $A$ and $B$ are determined by the microscopic structure of each wall. Then, we impose the partial slip boundary condition with slip length $b$ at $z=z_{0}$ and $z=L-z_{0}$, specifically, 
\begin{eqnarray}
\tilde{v}^x(\bm{r}) \Big|_{z=z_{0}} &=& b \frac{\partial \tilde{v}^x(\bm{r})}{\partial z} \Big|_{z=z_{0}},
\label{eq:partial slip boundary condition z=0}\\
 \tilde{v}^x(\bm{r}) \Big|_{z=L-z_{0}} &=& - b \frac{\partial \tilde{v}^x(\bm{r})}{\partial z} \Big|_{z=L-z_{0}},
\label{eq:partial slip boundary condition z=L} 
\end{eqnarray}
where $z_0$ is the hydrodynamic wall position.
Here, the slip length and the hydrodynamic wall position of walls $A$ and $B$ are equal because they have the same microscopic structure. With introducing the hydrodynamic wall position $z_0$, we also introduce the hydrodynamic system size $\mathcal{L}$ defined by
\begin{eqnarray}
\mathcal{L} = L - 2z_{0}.
\end{eqnarray}
In this section, we assume that $z_0$ is equal to $z_{\rm neq}$, and do not study the question (ii) (introduced in Sec.~\ref{subsec:Problem and strategy}). In Sec.~\ref{subsec:early-time behavior of gamma} and Appendix~\ref{subsec:discussion}, we will give some discussions.

\subsection{Explicit form of force autocorrelation function}
\label{subsec:explicit form of force autocorrelation function}
From LFH, the autocorrelation function of the force acting on the wall $\langle F(t)F(0)\rangle_{\rm eq}$ can be calculated. This was performed in Ref.~\cite{nakano2019statistical}. Here, we summarize the calculation results. The brief derivation is given in Appendix~\ref{sec:brief derivation of (42)}; see Secs. 7-10 in Ref.~\cite{nakano2019statistical} for details.

Because the fluid in equilibrium has time translational invariance, $\langle F(t)F(0)\rangle_{\rm eq}$ is expressed in the form
\begin{eqnarray}
\langle F(t)F(0)\rangle_{\rm eq} = \int \frac{d\omega}{2\pi} \langle |F(\omega)|^2\rangle_{\rm eq}e^{-i\omega t}.
\end{eqnarray}
From the linearity of the model, we can obtain the explicit expression of $\langle |F(\omega)|^2\rangle_{\rm eq}$:
\begin{widetext}
\begin{eqnarray}
\frac{\langle |F(\omega)|^2\rangle_{\rm eq}}{2\eta k_B T L_x L_y} &=& \Big|\frac{q}{\Delta}\Big|^2 \Big[\big(\frac{1}{q_R} + 6q_R b^2 \big)\sinh(2q_R \mathcal{L})  + 4\big(b + q_R^2 b^3\big) \cosh(2q_R \mathcal{L}) \nonumber \\
&+& \big(\frac{1}{q_R} - 6q_R b^2 \big) \sin(2q_R \mathcal{L}) + 4\big(b - q_R^2 b^3\big) \cos(2q_R \mathcal{L})
 \Big] \nonumber \\
 &+& \delta(0) \Big|\frac{q}{\Delta}\Big|^2 \Big[4 q_R b^3 \sinh(2q_R \mathcal{L}) - 4 q_R b^3 \sin(2q_R \mathcal{L}) \nonumber \\
 &+& 2 b^2\big(1+2q_R b^2 \big) \cosh(2q_R \mathcal{L}) + 2 b^2\big(1-2q_R b^2 \big) \cos(2q_R \mathcal{L}) + 4b^2 \Big] 
\label{eq:CAA expression: perfect form}
\end{eqnarray}
\end{widetext}
with
\begin{eqnarray}
\Delta = (1+qb)^2 e^{q\mathcal{L}} - (1-qb)^2 e^{-q\mathcal{L}} ,
\label{eq: def DELTA}
\end{eqnarray}
where $q$ is given by
\begin{eqnarray}
q = q_R - i q_R
\label{eq: def q}
\end{eqnarray}
with
\begin{eqnarray}
q_R = \sqrt{\frac{\omega \rho}{2 \eta}}.
\end{eqnarray}

Although the formal expression (\ref{eq:CAA expression: perfect form}) can be obtained by straightforward calculation, there is a difficulty. Specifically, the third and fourth lines of (\ref{eq:CAA expression: perfect form}) diverge  because these terms are proportional to $\delta(0)$. This divergence stems from the singularity of the stochastic flux at the same point (see eq. (\ref{eq:random stress tensor})). Here, we recall that the origin of the noise is the thermal motion of the molecules. Then, the properties of the stochastic flux (\ref{eq:random stress tensor}) appear as the result of coarse-graining  such molecular motion. Using this fact, the divergence of the delta function is regularized as follows. An infinitely small element in the coarse-graining description is assumed to be so large that it still contains a great number of molecules. The cutoff length $\xi_c$ is introduced as the minimum size of such a volume element; specifically, the infinitely small element in the coarse-graining description has volume $\xi_c^3$. Then, the properties of the stochastic flux (\ref{eq:random stress tensor}) implies that the correlation length is not equal to zero but is estimated as the cutoff length. Accordingly, the delta function in (\ref{eq:random stress tensor}), $\delta^3(\bm{r}-\bm{r}')$, is regularized as
\begin{eqnarray}
\delta^3(\bm{r}-\bm{r}') \simeq
\begin{cases}
 \xi_{\rm c}^{-3}, & {\rm for}  \ \ |\bm{r}-\bm{r}'| < \xi_{\rm c}, \\
0, & {\rm for} \ \ |\bm{r}-\bm{r}'| \geq \xi_{\rm c}.
\end{cases}
\end{eqnarray}
Similarly, by introducing the cutoff time $\tau_c$, we regularize the delta function $\delta(t-t')$ as
\begin{eqnarray}
\delta(t-t') \simeq
\begin{cases}
\tau_{\rm c}^{-1}, & {\rm for}  \ \ |t-t'| < \tau_{\rm c}, \\
0, & {\rm for} \ \ |t-t'| \geq \tau_{\rm c}.
\end{cases}
\end{eqnarray}

By the regularization of the stochastic flux, the singularity in (\ref{eq:CAA expression: perfect form}) is also regularized. Specifically, the delta function in (\ref{eq:CAA expression: perfect form}) is replaced by:
\begin{eqnarray}
\delta(0) \to \frac{1}{\xi_c}
\end{eqnarray}
Thus, the singularity in (\ref{eq:CAA expression: perfect form}) is understood to be the the cutoff-length dependence of the force autocorrelation function. In Ref.~\cite{nakano2019statistical}, we showed that such a cutoff-length dependence is approximately removed under the condition
\begin{eqnarray}
b \ll \xi_c \ll \mathcal{L}.
\end{eqnarray}
This condition is reasonable when the system size is fully macroscopic. However, in our simulation, this condition does not hold for two reasons. First, $\xi_c$ cannot be measured. Second, $b \ll \mathcal{L}$ does not hold. In particular, for Wall III, $b_{\rm neq}$ is over 6 times larger than the system size $\mathcal{L}$. Therefore, there is a problem with the validity of this condition. Nevertheless, in this paper, neglecting this problem, we assume that the cutoff-length dependence is not observed; the third and fourth lines of (\ref{eq:CAA expression: perfect form}) are neglected.

\subsection{Early-time behavior of $\gamma_{\rm LFH}(t)$}
\label{subsec:early-time behavior}
We study the Green--Kubo integral $\gamma_{\rm LFH}(t)$ of the force autocorrelation function calculated in LFH, which is given by
\begin{eqnarray}
\gamma_{\rm LFH}(t) =  \frac{1}{k_B T L_x L_y} \int_0^{t} ds \langle F(s)F(0) \rangle_{\rm eq}.
\end{eqnarray}
Note that $\gamma_{\rm LFH}(t)$ corresponds to (\ref{eq:gamma(t) in MD}) in the MD simulation. $\gamma_{\rm LFH}(t)$ is also expressed in the form
\begin{eqnarray}
\gamma_{\rm LFH}(t) = \frac{1}{k_B T L_x L_y} \int_0^{\infty} \frac{d\omega}{\pi} \langle |F(\omega)|^2\rangle_{\rm eq}\frac{\sin(\omega t)}{\omega}.
\label{eq:gamma t: CAAomega}
\end{eqnarray}
As explained in the previous subsection, $\langle |F(\omega)|^2\rangle_{\rm eq}$ (or $\langle F(s)F(0) \rangle_{\rm eq}$) is taken from (\ref{eq:CAA expression: perfect form}) where the third and fourth lines are neglected.

In this section, we focus on the behavior of $\gamma_{\rm LFH}(t)$ in the time region $t \ll \tau_{\rm macro}$, where $\tau_{\rm macro}$ is defined as the relaxation time of the velocity field. The behavior of $\gamma_{\rm LFH}(t)$ in this time region is obtained from $\langle |F(\omega)|^2\rangle_{\rm eq}$ in the frequency region $\omega \gg 2\pi/\tau_{\rm macro}$. Because the relaxation time of the velocity field is estimated as $\tau_{\rm macro} \sim \mathcal{L}^2 \rho/\eta$, the frequency region is rewritten as $\omega \gg 2\pi \eta/\mathcal{L}^2\rho$ or equivalently $q_R \gg \sqrt{\pi}/\mathcal{L}$. In this frequency region, $\langle |F(\omega)|^2\rangle_{\rm eq}$ is approximated by
\begin{eqnarray}
\frac{\langle |F(\omega)|^2\rangle_{\rm eq}}{2\eta k_B T L_x L_y} \simeq \frac{q_R(1+2q_Rb)}{1+2q_Rb+2q_R^2b^2}.
\label{eq:FF infty expansion}
\end{eqnarray}
Then, by substituting (\ref{eq:FF infty expansion}) into (\ref{eq:gamma t: CAAomega}), $\gamma_{\rm LFH}(t)$ in the time region $t \ll \tau_{\rm macro}$ is expressed as
\begin{eqnarray}
\gamma_{\rm LFH}(t) \simeq 2 \eta \int_0^{\infty} \frac{d\omega}{\pi} \frac{q_R(1+2q_Rb)}{1+2q_Rb+2q_R^2b^2}\frac{\sin(\omega t)}{\omega}.
\label{eq:gamma infty expansion}
\end{eqnarray}

From the expression (\ref{eq:gamma infty expansion}), we find two properties concerning the early-time behavior of $\gamma_{\rm LFH}(t)$. First, because (\ref{eq:gamma infty expansion}) does not contain the hydrodynamic system size $\mathcal{L}$, the early-time behavior of $\gamma_{\rm LFH}(t)$ is independent of the hydrodynamic system size $\mathcal{L}$. Second, $\gamma_{\rm LFH}(t)$ converges a finite value in the limit $t \to +0$, which is calculated as
\begin{eqnarray}
\lim_{t \to +0} \gamma_{\rm LFH}(t) &=& \frac{2\eta}{b} \lim_{t\to +0}\int_0^{\infty} \frac{d\omega}{\pi} \frac{\sin(\omega t)}{\omega}\nonumber \\[3pt]
&=& \frac{\eta}{b}.
\label{eq:Green--Kubo formula: microscopic fluctuating hydrodynamics}
\end{eqnarray}
See Ref.~\cite{nakano2019statistical} for more detailed arguments.

\section{Validity of LFH}
\label{sec:Validity of LFH description}
In this section, we demonstrate that LFH describes the fluctuations even at the molecular scale with high accuracy, although LFH was originally developed to describe the macroscopic fluctuations~\cite{landau1959course}. 


\subsection{Validity of LFH}
\label{subsec:validity of the LFH}
\begin{figure*}[ht]
\begin{center}
\subfigure[]{
\includegraphics[width=8cm]{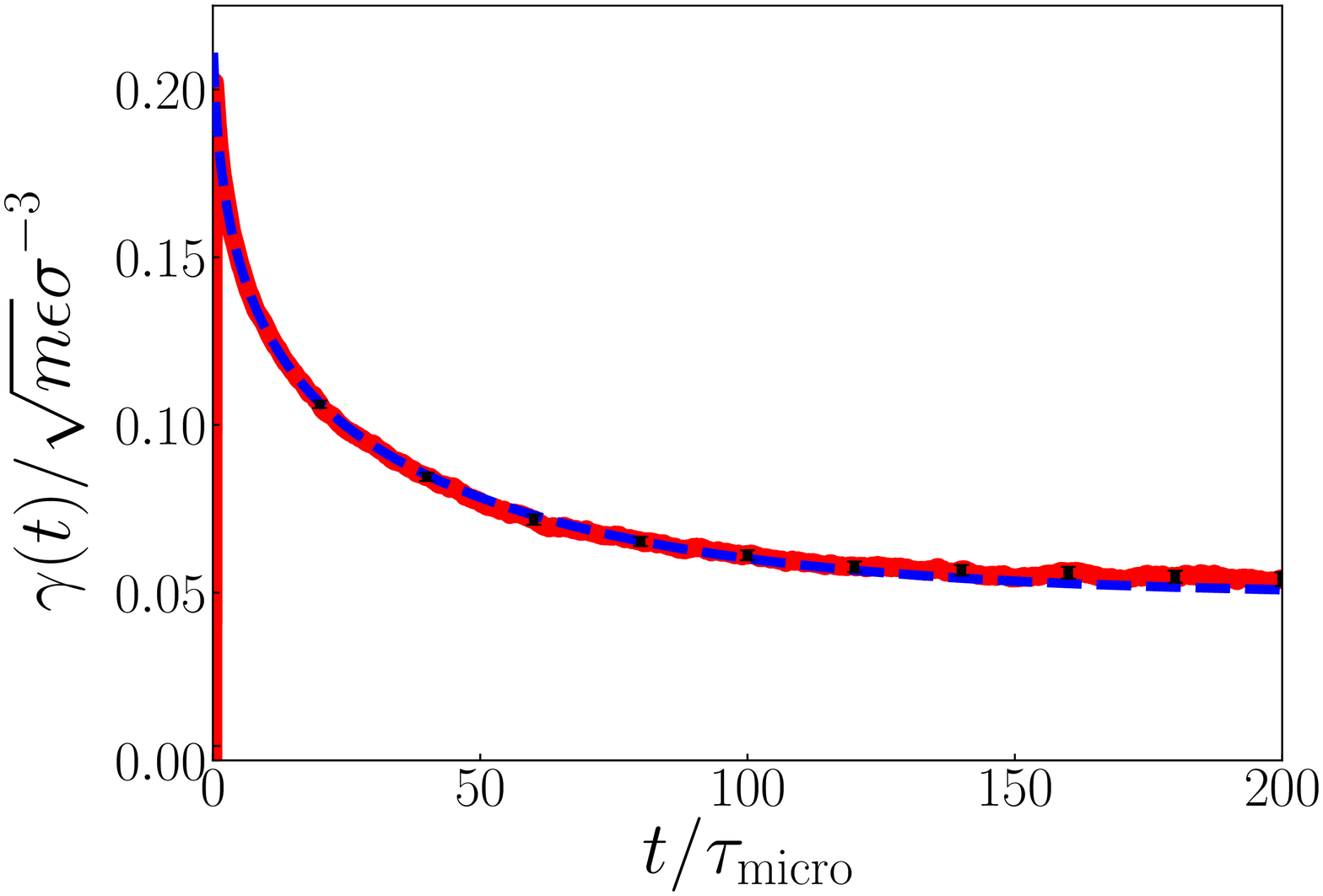}
\label{fig:cw0.80to200withfit}
} 
\subfigure[]{
\includegraphics[width=8cm]{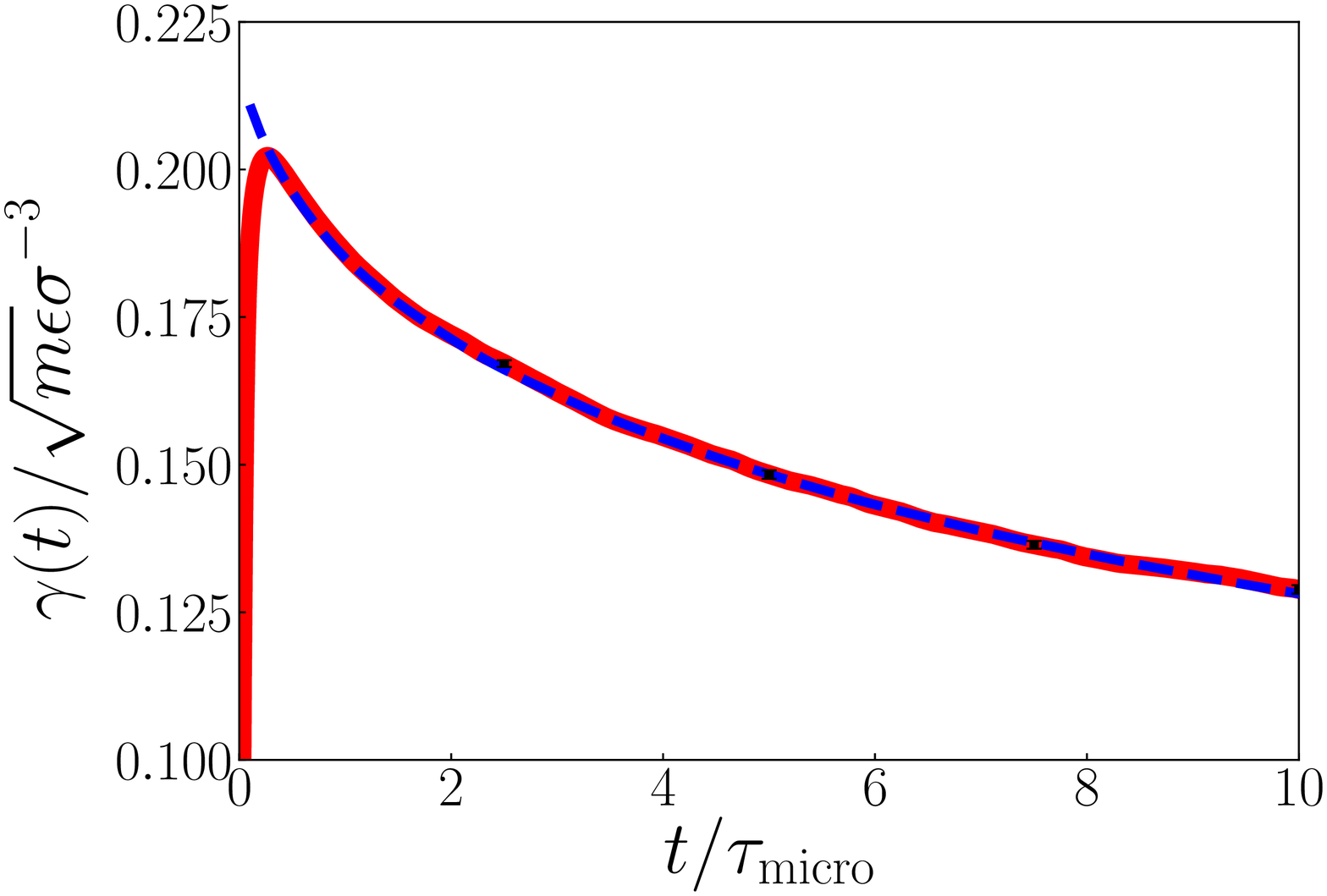}
\label{fig:cw0.80to20withfit}
}
\end{center}
\vspace{-0.5cm}
\caption{(Color online) $\gamma_{\rm LFH}(t)$ and $\gamma_{\rm MD}(t)$ for Wall I in time region (a) $[0,200\tau_{\rm micro}]$ and (b) $[0,10\tau_{\rm micro}]$. The red solid curves represent the simulation data (same as Fig.~\ref{fig:typical simulation result: aw0.6cw0.8}), and the error bars are the standard errors of means for each time for 24 different initial states. The blue broken curves represent LFH solutions with the best-fit parameters.}
\label{fig:typical comparison between EMD and LHD}
\end{figure*}
To examine whether LFH describes the MD simulation data, we compare $\gamma_{\rm LFH}(t)$ with $\gamma_{\rm MD}(t)$.
Fig.~\ref{fig:typical comparison between EMD and LHD} compares the simulation data with the LFH solution, which is obtained for Wall I. The left side (a) and right side (b) of Fig.~\ref{fig:typical comparison between EMD and LHD} present $\gamma_{\rm LFH}(t)$ and $\gamma_{\rm MD}(t)$ in the time regions $[0,200\tau_{\rm micro}]$ and $[0,10\tau_{\rm micro}]$, respectively. The slip length $b$ is determined by fitting the simulation data in the time region $[0,100\tau_{\rm micro}]$ to (\ref{eq:gamma t: CAAomega}). The fitting is performed by a non-linear least squares method (more precisely, scipy.optimize.curve\_fit in Python). When fitting $\gamma_{\rm MD}(t)$, we use the viscosity $\eta_{\rm neq}$ obtained in the NEMD simulation. The agreement is excellent when the slip length is $b_{\rm eq} = 7.73\sigma$. Here, the slip length obtained as the best-fit parameter is denoted by $b_{\rm eq}$.

We consider two characteristic time scales; one is that of the microscopic molecular motion, and the other is that of fluid relaxation. The typical microscopic time scale is given by $\tau_{\rm micro}$, while the relaxation time of the velocity field is estimated as $\tau_{\rm macro}\sim 176\tau_{\rm micro}$. From Fig.~\ref{fig:typical comparison between EMD and LHD}, the simulation data can be well fitted by the LFH solution from the $\tau_{\rm micro}$-scale to the $\tau_{\rm macro}$-scale. In particular, the excellent agreement in the time region $[0,10\tau_{\rm micro}]$ (see Fig.~\ref{fig:cw0.80to20withfit}) may be a surprising result because the fluctuating hydrodynamics was originally developed to describe relaxation processes of macroscopic fluids.

\begin{figure}[ht]
\begin{center}
\includegraphics[width=4cm]{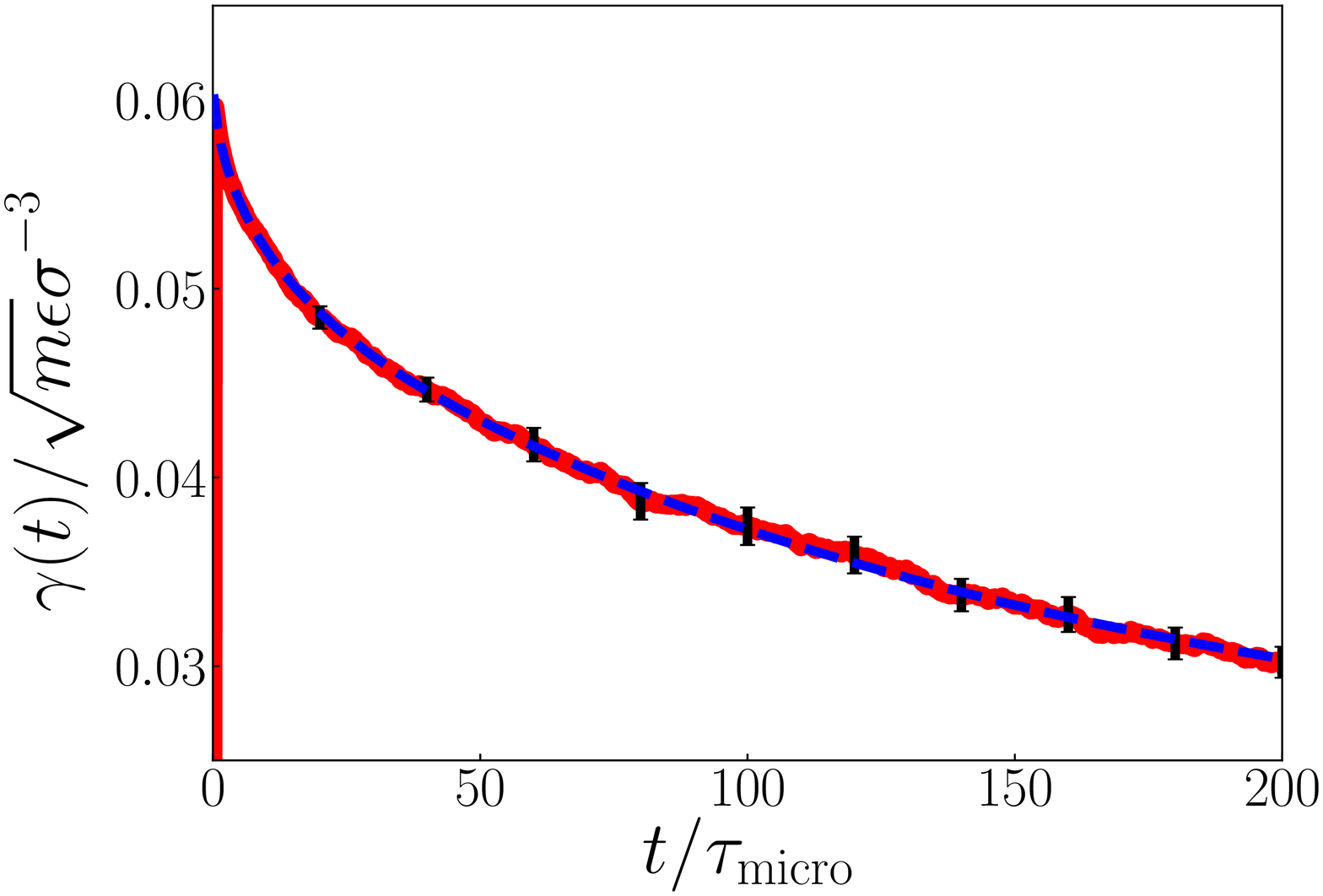} 
\includegraphics[width=4cm]{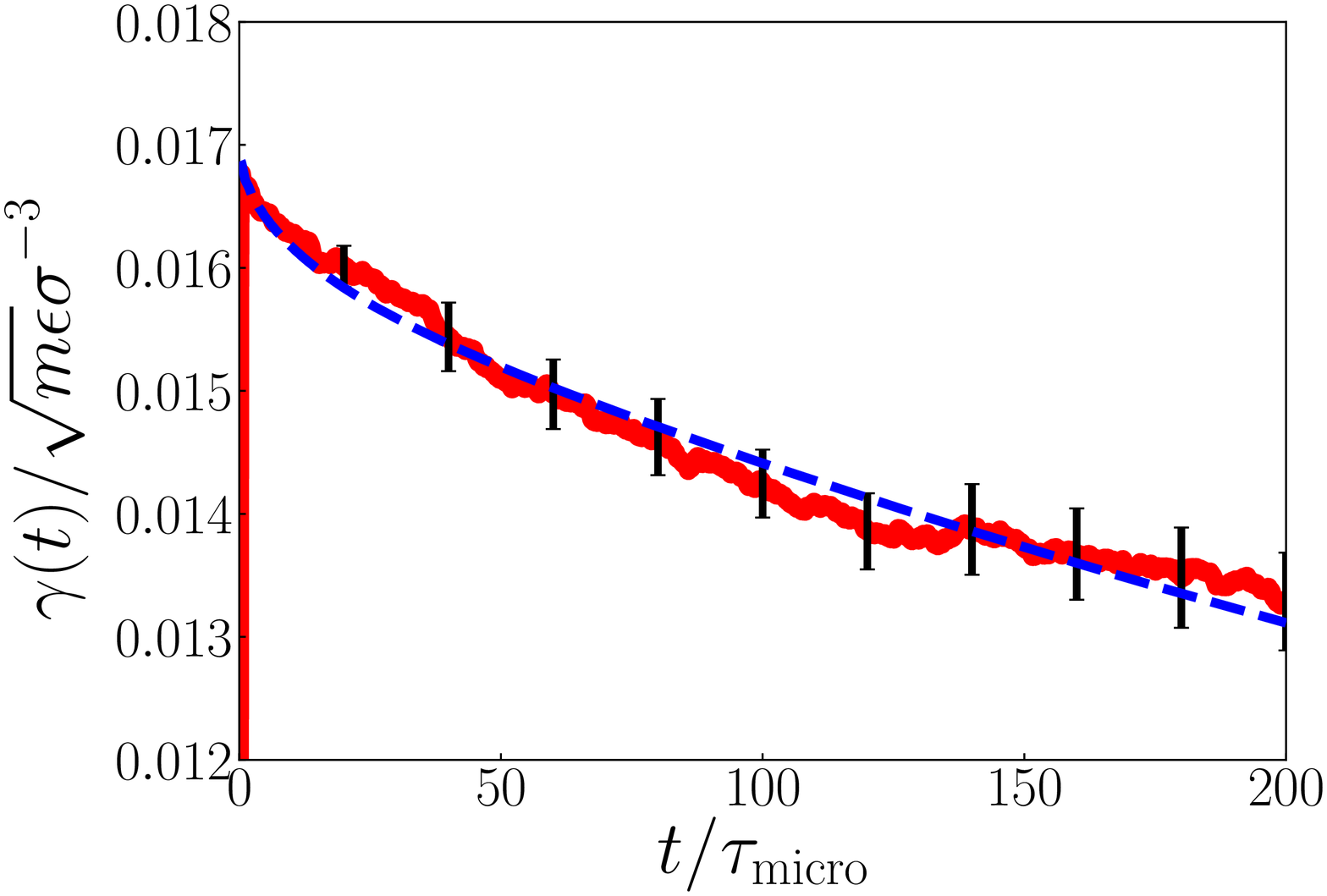} 
\end{center}
\vspace{-0.5cm}
\caption{(Color online) Same as Fig.~\ref{fig:typical comparison between EMD and LHD}, but with different walls. Left: Wall II. Right: Wall III.}
\label{fig:another comparison between EMD and LHD}
\end{figure}

This excellent agreement between the simulation data and the LFH solution is found for all the atomically smooth surfaces examined. Figure~\ref{fig:another comparison between EMD and LHD} presents the behavior of $\gamma_{\rm MD}(t)$ and $\gamma_{\rm LFH}(t)$ in the time region $[0,200\tau_{\rm micro}]$ for Walls II and III. Therefore, we conclude that LFH is the accurate model to describe the behavior of $\gamma_{\rm MD}(t)$.

From another aspect, this excellent agreement indicates the validity of $(\rho,\eta)$ that we have used when comparing the LFH solution and simulation data. In particular, as explained in Sec.~\ref{subsec:Linearized fluctuating hydrodynamics}, we have neglected the inhomogeneous density field near the walls and have used the density in the bulk region. Also, the viscosity $\eta$ is assumed to be independent of $z$ even near the walls. This assumption implies that the bulk constitutive equation holds near the walls. The validity of these assumptions is also confirmed.

\subsection{New equilibrium measurement method}
\label{subsec:proposal of new equilibrium measurement method}
By comparing the LFH solution and the simulation data, we confirm the validity of LFH introduced in Sec.~\ref{subsec:Linearized fluctuating hydrodynamics}. Accordingly, the slip length $b_{\rm eq}$, which is obtained as the best-fit parameter, turns out to characterize the partial slip boundary condition imposed in LFH. On the other hand, in the non-equilibrium measurement method, we first investigate the validity of the bulk constitutive equation with the partial slip boundary condition. By confirming this validity, we find that the slip length $b_{\rm neq}$ obtained in the NEMD simulation characterizes the partial slip boundary condition imposed in the deterministic hydrodynamics.

Clearly, it is reasonable to conjecture that the partial slip boundary condition imposed in LFH is identical to the one imposed in the deterministic hydrodynamics. Then, in order to confirm this conjecture, we compare $b_{\rm eq}$ with $b_{\rm neq}$. In table~\ref{tab:comparison of slip length}, we summarize $b_{\rm neq}$, $b_{\rm peak}$ and $b_{\rm eq}$ for the three types of the walls.
\begin{table*}[]
\begin{center}
\begin{tabularx}{160mm}{C||CCC}\hline
Wall & $b_{\rm neq}/\sigma$ & $b_{\rm peak}/\sigma$ & $b_{\rm eq}/\sigma$ \\ \hline\hline
I \ \  \modHN{$(c_{AF}=0.8)$} & $8.08\pm0.14$ & $8.71\pm0.05$ & $7.73\pm0.08$\\
II \ \  \modHN{$(c_{AF}=0.4)$} & $30.8\pm0.3$ & $29.8\pm0.3$ & $28.7\pm0.5$ \\
III \ \  \modHN{$(c_{AF}=0.0)$} & $121.8\pm1.7$ & $107.7\pm1.8$ & $106.6\pm3.1$ \\\hline\hline
\end{tabularx}
\caption{Comparison of slip lengths calculated from three different methods. Column 1 gives the wall parameters. $b_{\rm neq}$ is the slip length calculated from the NEMD simulation (Sec.~\ref{subsec:Non-equilibrium molecular dynamics simulation}). $b_{\rm peak}$ is the slip length estimated from BB's relation (\ref{eq:BB's relation revisted2}) (Sec.~\ref{sec:Review of BB's formula}). $b_{\rm eq}$ is the slip length measured by the new equilibrium measurement method proposed in Sec.~\ref{subsec:proposal of new equilibrium measurement method}. For errors, see each section. }
\label{tab:comparison of slip length}
\end{center}
\end{table*}
The error given for $b_{\rm eq}$ is the standard deviation, which is estimated from the measurement error of the simulation data for different 24 initial states and that of the viscosity $\eta_{\rm neq}$. The deviations between $b_{\rm eq}$ and $b_{\rm neq}$ remain within $15\%$. Then, by allowing for these deviations, we conclude that $b_{\rm eq}$ approximately coincides with $b_{\rm neq}$. 

Conversely, when we allow that $b_{\rm eq}$ coincides with $b_{\rm beq}$, we can estimate $b_{\rm neq}$ from the EMD simulation, which procedure is summarized as follows. 
\begin{enumerate}
\item measurement of $\eta$ in advance
\item measurement of $\gamma_{\rm MD}(t)$ from the EMD simulation
\item fitting of $\gamma_{\rm MD}(t)$ to the LFH solution $\gamma_{\rm LFH}(t)$
\end{enumerate}
If the excellent agreement is found, the slip length is obtained as the best-fit parameter.

Here, we note that our method requires the value of the viscosity $\eta$ in advance. In this paper, we used the viscosity $\eta_{\rm neq}$ obtained in the NEMD simulation. However, because the viscosity $\eta$ is also obtained from the EMD simulation~\cite{hansen2013theory}, our method can be closed in the EMD simulation.

Also, we recall that LFH reproduces the simulation data down to the $\tau_{\rm micro}$-scale. From this fact, we can estimate the slip length using only the simulation data in the early-time region. Actually, although in Sec.~\ref{subsec:validity of the LFH} the fitting is performed for $\gamma_{\rm MD}(t)$ in $[0,100\tau_{\rm micro}]$, the almost same results are obtained by using the simulation data in $[0,20\tau_{\rm micro}]$. Accordingly, our method is performed at a relatively low computational cost.

We comment on the differences between our equilibrium measurement method and BB's relation. First, whereas BB's relation uses the instant information (specifically $t=\tau_0$) in order to obtain the slip length, our method uses the information of $\gamma_{\rm MD}(t)$ over the finite time region. Then, from the viewpoint of extending the time region used to calculate the slip length, our method is interpreted as an extension to BB's relation. Second, in order to obtain the friction coefficient $\lambda$, the relation (\ref{eq:BB's relation revisted1}) does not require the value of viscosity $\eta$. On the other hand, our method always requires the viscosity $\eta$ because the LFH solution (\ref{eq:CAA expression: perfect form}) contains the viscosity $\eta$ in all time region. This is a crucial difference.

\subsection{Early-time behavior of $\gamma_{\rm MD}(t)$}
\label{subsec:early-time behavior of gamma}
In Sec.~\ref{subsec:early-time behavior}, we derived from the full LFH solution that the early-time behavior of $\gamma_{\rm LFH}(t)$ is independent of the system size $\mathcal{L}$ or equivalently the hydrodynamic wall position $z_0$. In this subsection, we show that this result is consistent with the EMD simulation result.

\begin{figure}[t]
\centering
\includegraphics[width=8cm]{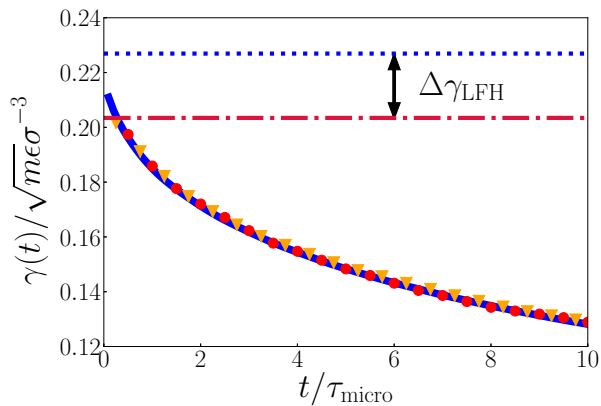} 
\caption{(Color online) System size dependence of $\gamma_{\rm MD}(t)$. The wall parameter is \modHN{$c_{AF}=0.8$} (Wall I), which is the same as Fig.~\ref{fig:typical simulation result: aw0.6cw0.8}. The blue solid curve is $\gamma_{\rm LFH}(t)$ with the best-fit parameter. The red circles and orange triangles are simulation data for $L=20.0\sigma$ and $L=30.0\sigma$, respectively. The red dash-dot line gives the value of the first peak of $\gamma_{\rm MD}(t)$ and the blue dot line gives the extrapolated value of $\gamma_{\rm LFH}(t)$ to $t \to +0$. $\Delta_{\rm LFH}$ is defined by (\ref{eq:deviation of delta gamma})}
\label{fig:System Size Dependence of EMD}
\end{figure}

For this purpose, we perform an additional EMD simulation with Wall I and $L=30.0\sigma$. The result is presented as the orange triangles in Fig.~\ref{fig:System Size Dependence of EMD}. For comparison, the simulation result with $L=20.0\sigma$ is displayed as the red circles, which corresponds to the red solid curve in Fig.~\ref{fig:cw0.80to20withfit}. In addition, the blue solid curve in Fig.\ref{fig:System Size Dependence of EMD} represents the LFH solution with $b_{\rm eq}=7.73\sigma$, which corresponds to the blue broken curve in Fig.~\ref{fig:cw0.80to20withfit}. From these graphs, we conclude that the decay of $\gamma_{\rm MD}(t)$ from the first peak is independent of $L$ (or equivalently $z_0$). \modHN{This behavior is consistent with that recently  reported by J. A. de la Torre \textit{et al.}~\cite{PhysRevE.99.022126,PhysRevLett.123.264501}, where they claimed that $\gamma_{\rm MD}(t)$ does not exhibit the plateau region even in the thermodynamic limit. This claim implies that the early-time behavior of $\gamma_{\rm MD}(t)$ is independent of $L$.}

By combining this result with the fact that the slip length can be estimated from the early-time behavior of $\gamma_{\rm MD}(t)$ (see Sec.~\ref{subsec:proposal of new equilibrium measurement method}), we obtain two practical notes concerning the new equilibrium measurement method of the slip length. First, the measurement accuracy of the slip length is independent of the system size $L$ because the simulation result is independent of the system size $L$. Second, the slip length $b_{\rm eq}$ is determined from the EMD simulation result independently of the hydrodynamic position $z_0$. In Sec.~\ref{subsec:Linearized fluctuating hydrodynamics}, we assumed that the hydrodynamic wall position $z_0$ is equal to $z_{\rm neq}$. From the second note, we find that this assumption is not necessary to measuring the slip length $b_{\rm eq}$.

\section{BB's relation revisited}
\label{sec:BB's relation revisited}
Combining LFH and the EMD simulation yields the new theoretical interpretation of BB's relation. In this section, we explain it.

Let us consider the behavior of $\gamma_{\rm MD}(t)$ and $\gamma_{\rm LFH}(t)$ in the $\tau_{\rm micro}$-scale. As shown in Sec.~\ref{subsec:validity of the LFH}, the simulation data is well fitted by the LFH solution even in the $\tau_{\rm micro}$-scale (see Fig.~\ref{fig:cw0.80to20withfit}). Then, we have
\begin{eqnarray}
\gamma_{\rm MD}(\tau_0) = \gamma_{\rm LFH}(\tau_0).
\label{eq:gammaMD and gammaLFH}
\end{eqnarray}
However, the behavior of $\gamma_{\rm MD}(t)$ in the time region $[0,\tau_0]$ is different from that of $\gamma_{\rm LFH}(t)$. The MD simulation shows that $\gamma_{\rm MD}(t)$ grows from $t=0$ to $t=\tau_0$ and decays after that (see Fig.~\ref{fig:cw0.80to20withfit}). Whereas, as shown in Sec.~\ref{subsec:early-time behavior}, $\gamma_{\rm LFH}(t)$ does not grow in the time region $[0,\tau_0]$; instead, it has the finite value (\ref{eq:Green--Kubo formula: microscopic fluctuating hydrodynamics}) at $t=0$. Then, we define the deviation between $\gamma_{\rm LFH}(0)$ and $\gamma_{\rm LFH}(\tau_0)$ as
\begin{eqnarray}
\hspace{-0.5cm}\Delta \gamma_{\rm LFH} &\equiv& \gamma_{\rm LFH}(0) - \gamma_{\rm LFH}(\tau_0) \nonumber \\[3pt]
&\simeq& 2 \frac{\eta}{b_{\rm eq}} \int_0^{\infty} \frac{d\omega}{\pi} \frac{ 1+ q_R b_{\rm eq}}{1+2q_Rb_{\rm eq}+2q_R^2b_{\rm eq}^2}\frac{\sin(\omega \tau_0)}{\omega},\hspace{0.2cm}
\label{eq:deviation of delta gamma}
\end{eqnarray}
where the second line was obtained by substituting (\ref{eq:gamma infty expansion}) and (\ref{eq:Green--Kubo formula: microscopic fluctuating hydrodynamics}) into the first line. By using (\ref{eq:Green--Kubo formula: microscopic fluctuating hydrodynamics}), (\ref{eq:gammaMD and gammaLFH}) and (\ref{eq:deviation of delta gamma}), we obtain
\begin{eqnarray}
\frac{\eta}{b_{\rm eq}} &=& \gamma_{\rm MD}(\tau_0) + \Delta \gamma_{\rm LFH} .
\label{eq:deviation of gamma}
\end{eqnarray}

Using this equation (\ref{eq:deviation of gamma}), we propose the new theoretical interpretation of BB's relation (\ref{eq:BB's relation revisted2}). By comparing (\ref{eq:deviation of gamma}) with BB's relation (\ref{eq:BB's relation revisted2}), we find that $\Delta\gamma_{\rm LFH}$ gives the remainder term of BB's relation. More precisely, we can identify the accuracy of BB's relation from this value. BB's relation provides the reasonable estimation of the slip length $b_{\rm eq}$ when 
\begin{eqnarray}
\frac{\Delta \gamma_{\rm LFH}}{\gamma_{\rm LFH}(\tau_0)} \ll 1
\end{eqnarray}
holds.

As a example, in Fig.~\ref{fig:System Size Dependence of EMD}, we depict $\Delta \gamma_{\rm LFH}$ for Wall I, where we have
 \begin{eqnarray}
\frac{\Delta \gamma_{\rm LFH}}{\gamma_{\rm LFH}(\tau_0)} \simeq 0.129.
\end{eqnarray}
Allowing for this deviation, we can obtain the estimation of the slip length $b_{\rm eq}$ from BB's relation.

There are two notes on our approach for BB's relation. First, our argument relates $b_{\rm peak}$ to not $b_{\rm neq}$ to $b_{\rm eq}$. In order to relate $b_{\rm peak}$ to $b_{\rm neq}$, our approach requires the additional assumption that $b_{\rm eq}$ is equal to $b_{\rm neq}$. Second, the deviation between $b_{\rm eq}$ and $b_{\rm peak}$ always exists because $\Delta \gamma_{\rm LFH}$ is not equal to zero by definition.

\begin{figure*}[ht]
\centering
\includegraphics[width=15cm]{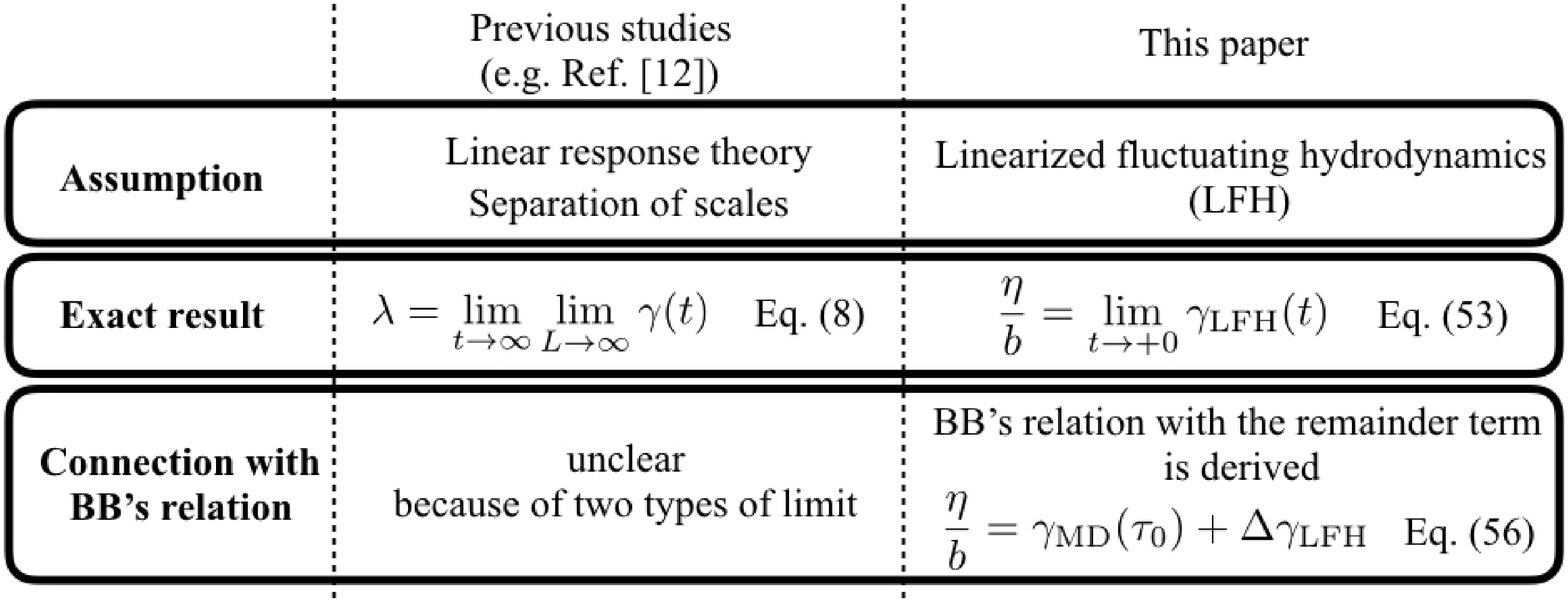} 
\caption{Diagram illustrating the comparison between the approach to BB's relation in the previous studies~\cite{bocquet1994hydrodynamic,fuchs2002statistical,kobryn2008molecular,bocquet2013green,chen2015determining,camargo2019boundary} and that in this paper.}
\label{fig:diagram illustrating}
\end{figure*}

In Fig.~\ref{fig:diagram illustrating}, we summarize the difference between our approach for BB's relation and that in the previous studies. The crucial difference is the starting point. Whereas the previous studies started with assuming the validity of the linear response theory and the separation of scales (see Introduction), we start with assuming that LFH accurately describes the fluctuation of the confined fluid even in the $\tau_{\rm micro}$-scale. While our assumption is easily verified by numerical simulations as shown in Sec.~\ref{sec:Validity of LFH description}, it may be difficult to verify the assumption imposed in the previous studies because of the two types of limit.

\section{Discussions}
\label{sec:Discussion}
In this paper, we demonstrated that for atomically smooth walls, LFH accurately reproduces the force autocorrelation function from the time scale of molecular motion to that of fluid relaxation. As a result, LFH is a useful starting point to analyze the behavior of the force autocorrelation function. Furthermore, the slip length obtained as the best-fit parameter is in excellent agreement with that obtained from the non-equilibrium measurement. This fact yields the new equilibrium measurement method.

We also showed that LFH provides the new theoretical interpretation of BB's relation (\ref{eq:BB's relation revisted2}). The starting point for our argument was (\ref{eq:gammaMD and gammaLFH}), which implies that the LFH description is valid at the microscopic scale. Then, our argument can be applied to more realistic and complicated walls, beyond atomically smooth walls, as long as $\gamma_{\rm MD}(t)$ can be well-fitted by the LFH solution.

We stress that our equilibrium measurement method is based on the approximate solution calculated from LFH. That is, throughout this paper, our analysis was performed by neglecting the third and fourth line of (\ref{eq:CAA expression: perfect form}). This approximation is theoretically justified when we assume the condition $b \ll \xi_c \ll \mathcal{L}$~\cite{nakano2019statistical}. In this paper, we demonstrated that this approximation can be used even for the systems with $b \simeq \mathcal{L}$. The theoretical verification of this result is a future task. Furthermore, from the applied viewpoint, the interesting situations are when the condition $b\gg L$ holds~\cite{falk2010molecular,kannam2013fast,secchi2016massive}. It is interesting to examine such systems numerically and theoretically.

Recently, some different equilibrium measurement methods for the slip length were proposed~\cite{petravic2007equilibrium,hansen2011prediction,kumar2012slip,huang2014green,oga2019green,sokhan2008slip,sam2018prediction}. They basically extract the friction coefficient $\lambda$ or the slip length $b$ from the autocorrelation function of the force acting on the wall, $\langle F(t)F(0) \rangle$, that of the fluid velocity near the wall, $\langle v^x(z,t) v^x(z,0)\rangle$, or the cross-correlation function, $\langle F(t) v^x(z,0)\rangle$. In the previous paper~\cite{nakano2019statistical}, we showed that the validity of the methods in Refs.~\cite{petravic2007equilibrium,huang2014green,hansen2011prediction} can be understood from LFH. However, it is desirable to compare their methods and our one from the viewpoint of the computational cost and the measurement accuracy.

\modHN{
There are two especially interesting studies. First, J. A. de la Torre \textit{et al.} proposed a new method to estimate $\tau_0$ in BB's relation (\ref{eq: expression of slip length: bocquet and barrat})~\cite{PhysRevE.99.022126,PhysRevLett.123.264501}. Their method is given by the following steps; (i) we divide the fluid confined between the two parallel walls into $N_{\rm bin}$ bins; (ii) we calculate the correlation matrix $C_{\mu \nu}(t)=\langle \hat{g}_{\mu}\hat{g}_{\nu}\rangle$ where $\hat{g}_{\mu}$ is the total momentum density in bin $\mu$; (iii) for larger bin width, all modes of $C_{\mu \nu}(t)$ exhibit the exponentially decay. $\tau_0$ is estimated as the maximum decay time for this bin width. Although $\tau_0$ obtained from their method is slightly different from the first peak of $\gamma(t)$, we can repeat the argument in Sec.~\ref{sec:BB's relation revisited}, which leads to the result that the slip length $b_{\rm eq}$ is slightly different from that obtained by J. A. de la Torre \textit{et al.}. Therefore, it is an interesting problem how the theory of J. A. de la Torre \textit{et al.} is connected with the fluctuating hydrodynamics.

Second,} H. Oga et al.~\cite{oga2019green} introduced another model to reproduce the early-time behavior of $\gamma_{\rm MD}(t)$. It is given by the Langevin equation, which describes the behavior of a coarse-grained system involving a few atoms thickness of fluid adjacent to the wall. Here, we comment on the differences between their phenomenological model and LFH. The crucial difference is that their model does not use information of the bulk region and contains three parameters related to the properties of fluids adjacent to the wall, specifically, the mass of focused fluid layer, characteristic time of the viscoelastic motion of that, and friction coefficient. On the other hand, LFH contains three parameters $(\eta,\rho,b)$ related to the fluid in the bulk region. In order to more deeply understand the fluctuating dynamics of the fluid adjacent to the wall, it is desirable to elucidate the relationship between these parameters and then to establish the relationship between their model and LFH.

Finally, we comment on our result from the aspect of the validity of the continuum description in extremely small systems. Previous studies mainly examined the validity of the deterministic hydrodynamics, i.e. the limit of the bulk constitutive equation, the partial slip boundary condition, and the system-size dependence of the viscosity and the slip length~\cite{bocquet2010nanofluidics}. On the other hand, our study focuses on the fluctuating hydrodynamics. Whereas our result demonstrated that LFH holds in the system with about $20$-atoms thickness, it is known that for a certain system the deterministic hydrodynamics holds even when the system width is equal to $3$-atoms layer~\cite{bocquet2010nanofluidics}. Therefore, it is one of the important challenges to elucidate the limit of LFH, particularly, whether LFH holds in the system with a-few-atoms thickness.

This also involves the theoretical problem of non-equilibrium statistical mechanics, specifically, the derivation of the fluctuating hydrodynamics from the underlying microscopic dynamics. So far, the fluctuating hydrodynamic has been derived using various methods~\cite{zwanzig1961memory,kawasaki1973simple,zubarev1996statistical}. These derivations show that fluctuating hydrodynamics is rigorously valid in the thermodynamic limit. However, our simulation results are clearly beyond the scope of previous derivations. There are few statistical mechanical tools to handle such a problem, and a new concept would be necessary. 

\onecolumngrid
\begin{acknowledgments}
The present study was supported by KAKENHI (Nos. 17H01148, 19H05496, 19H05795).
\end{acknowledgments}

\appendix
\section{Brief derivation of (\ref{eq:CAA expression: perfect form})}
\label{sec:brief derivation of (42)}
In this Appendix, we give a brief derivation of (\ref{eq:CAA expression: perfect form}). See Secs. 7-10 in Ref.~\cite{nakano2019statistical} for the detailed argument.

In the continuum description, the force acting on wall $A$, $F(t)$ is defined as
\begin{eqnarray}
F(t) &=& \int_{z=z_0+\xi_{c}} \hspace{-0.15cm} dx dy~\tilde{J}^{xz}(\bm{r},t)  \nonumber \\[3pt]
&=& - \eta \partial_{z} \tilde{\mathcal{V}}^{x}(z_0+\xi_{c},t) + \tilde{\mathcal{S}}^{xz}(z_0+\xi_{c},t) ,
\label{eq:explicit form of FA}
\end{eqnarray} 
where 
\begin{eqnarray}
\tilde{\mathcal{V}}^{x}(z,t) &=& \int_{z} dx dy~\tilde{v}^{x}(\bm{r},t),
\label{eq:def of bar(v)}\\
\tilde{\mathcal{S}}^{xz}(z,t) &=& \int_{z} dx dy~\tilde{s}^{xz}(\bm{r},t).
\label{eq:def of bar(s)}
\end{eqnarray}
Then, the force autocorrelation function in the frequency domain is expressed as
\begin{eqnarray}
 \langle |F(\omega)|^2\rangle_{\rm eq} &=& \eta^2 \Big\langle \partial_{z} \tilde{\mathcal{V}}^{x}(z_0+\xi_{c},\omega) \partial_{z} \tilde{\mathcal{V}}^{x}(z_0+\xi_{c},-\omega) \Big\rangle_{\rm eq} + \Big\langle \tilde{\mathcal{S}}^{xz}(z_0+\xi_{c},\omega) \tilde{\mathcal{S}}^{xz}(z_0+\xi_{c},-\omega) \Big \rangle_{\rm eq} \nonumber \\[3pt]
&-& \eta \Big\langle \partial_{z} \tilde{\mathcal{V}}^{x}(z_0+\xi_{c},\omega)\tilde{\mathcal{S}}^{xz}(z_0+\xi_{c},-\omega) \Big\rangle_{\rm eq} - \eta \Big\langle \partial_{z}\tilde{\mathcal{V}}^{x}(z_0+\xi_{c},-\omega) \tilde{\mathcal{S}}^{xz}(z_0+\xi_{c},\omega) \Big\rangle_{\rm eq} .
\label{eq:CAA: frequency domain}
\end{eqnarray}
Because our model is linear, $\tilde{\mathcal{V}}^{x}(z,\omega)$ is expressed using $\tilde{\mathcal{S}}^{xz}(z,\omega)$ as
\begin{eqnarray}
\tilde{\mathcal{V}}^{x}(z,\omega) = \int_{z_0}^{\mathcal{L}+z_0} dz' G(z,z',\omega) \partial_{z'} \tilde{\mathcal{S}}^{xz}(z',\omega),
\label{eq:vx in terms of Green function}
\end{eqnarray}
where $G(z,z',\omega)$ is the Green function. The Green function $G(z,z',\omega)$ is calculated by substituting (\ref{eq:vx in terms of Green function}) into (\ref{eq:Navier-Stokes equation subjected to a Gaussian random stress tensor}), (\ref{eq:partial slip boundary condition z=0}) and (\ref{eq:partial slip boundary condition z=L}). After straightforward calculation, we obtain
\begin{eqnarray}
G(z,z';\omega) = g_1(z') e^{-q (z-z_0)} + g_2(z') e^{q(z-\mathcal{L}-z_0)} + \frac{1}{2\eta q}e^{-q|z-z'|}
\label{eq:Green function: explicit form}
\end{eqnarray}
with
\begin{eqnarray}
g_1(z') &=& - \frac{1}{2\eta q \Delta} \Big((1-q^2 b^2) e^{q(\mathcal{L}+z_0-z')} - (1-qb)^2 e^{-q(\mathcal{L}+z_0-z')} \Big),
\label{eq:g1}\\
g_2(z') &=& - \frac{1}{2\eta q \Delta} \Big((1-q^2b^2) e^{q(z'-z_0)} - (1-qb)^2 e^{-q(z'-z_0)} \Big),
\label{eq:g2}
\end{eqnarray}
where $\Delta$ and $q$ are given by (\ref{eq: def DELTA}) and (\ref{eq: def q}), respectively. By substituting (\ref{eq:vx in terms of Green function}) and (\ref{eq:Green function: explicit form}) into (\ref{eq:CAA: frequency domain}), we obtain (\ref{eq:CAA expression: perfect form}).

\vspace{0.5cm}
\twocolumngrid
\vspace{0.5cm}
\section{Equilibrium measurement of hydrodynamic wall position}
\label{subsec:discussion}
As shown in Sec~\ref{subsec:early-time behavior of gamma}, when we use $\gamma_{\rm MD}(t)$ in the early-time region in order to estimate the slip length $b_{\rm eq}$, our equilibrium measurement method does not need the value of the hydrodynamic wall position $z_0$. However, in the main text, we did not give the method to calculate the hydrodynamic wall position $z_0$ from the EMD simulation. In this Appendix, from the theoretical viewpoint, we explain that the hydrodynamic wall position $z_0$ can be measured by using $\gamma_{\rm MD}(t)$ in the late-time region.

In principle, $z_0$ can be measured by using $(b,z_0)$ as the fitting parameters ($(\eta,\rho)$ are input parameters). Because the full LFH solution contains $z_0$ in the form $\mathcal{L}=L-2z_0$, the measurement of $z_0$ is equivalent to that of $\mathcal{L}$. By noting that $\gamma_{\rm LFH}(t)$ is related to the force autocorrelation function of wall $A$, we find that the $\mathcal{L}$-dependence of $\gamma_{\rm LFH}(t)$ is mainly observed in the time region $t\gtrsim \tau_{\rm macro}$.

In particular, as shown in Ref.~\cite{nakano2019statistical}, we have
\begin{eqnarray}
\lim_{t \to\infty} \gamma_{\rm LFH}(t) = \frac{\eta}{\mathcal{L}+2b_{\rm eq}},
\label{eq:Petravic and Harrowell}
\end{eqnarray}
where more precisely $t\to \infty$ means the limit $t\gg \tau_{\rm macro}$. Then, by measuring the convergence value of $\gamma_{\rm MD}(t)$ in the EMD simulation, we obtain $\mathcal{L}+2b_{\rm eq}$. Because $b_{\rm eq}$ is obtained from the early-time behavior of $\gamma_{\rm MD}(t)$, by combining these equilibrium measurements, we obtain $\mathcal{L}$ or $z_0$.

However, this equilibrium measurement method requires high computational cost. Actually, for Wall I, $\gamma_{\rm MD}(t)$ converges at $t=500\tau_{\rm micro}$ and the value is $\gamma_{\rm MD}(500\tau_{\rm micro}) = 0.0524\pm0.0170$, where the error represents the standard deviation from $24$ independent simulations. Then, from (\ref{eq:Petravic and Harrowell}), the hydrodynamic system size is calculated as $\mathcal{L}=18.1\pm 10.9$. The error is rather larger than that of the slip length (see table~\ref{tab:comparison of slip length}). If we neglect this error, the hydrodynamic wall position is calculated as $z_0=1.05$; this value is close to the value of $z_{\rm neq}$ obtained in the NEMD simulation. However, more detailed discussion taken the error into account remains as the future work. We here comment on why the measurement error of $\mathcal{L}$ is much larger than that of $b_{\rm eq}$. First, in order to obtain $\gamma_{\rm MD}(t)$ in the time region $t\gg \tau_{\rm macro}$ from the EMD simulation,  we must perform the simulation with the sufficiently large $\Delta T_{\rm obs}$. Second, because the convergence value of $\gamma_{\rm MD}(t)$ is rather small, the slight error leads to larger error in the hydrodynamic wall position $\mathcal{L}$.

%

\end{document}